\documentclass[12pt]{article}
     \usepackage{amsmath}
  \usepackage{graphicx}
  \newlength{\abstractwidth}
  \renewcommand{\thanks}[1]{\footnote{#1}} 

  \newcommand{\be}{\begin{equation}}
  \newcommand{\bea}{\begin{eqnarray}}
  \newcommand{\eea}{\end{eqnarray}}
  \newcommand{\beq}{\begin{equation}}
  \newcommand{\ee}{\end{equation}}
  \newcommand{\eeq}{\end{equation}}

  \def\ba{\begin{eqnarray}}
  \def\ea{\end{eqnarray}}

  \def\12{{1 \over 2}}

 \def\simleq{\; \raise0.3ex\hbox{$<$\kern-0.75em
      \raise-1.1ex\hbox{$\sim$}}\; }
 \def\simgeq{\; \raise0.3ex\hbox{$>$\kern-0.75em
      \raise-1.1ex\hbox{$\sim$}}\; }

\def\O2{\Omega_2}

\def\bi{\begin{itemize}}
  \def\ei{\end{itemize}}

\def\W{$\Omega$}
\def\W'{$\Omega$}

\def\V{\Omega}
\def\V'{\Omega}

\def\nref#1{ (\ref{#1})}

\begin{document}

\begin{titlepage}
  \bigskip

  \bigskip\bigskip

  \bigskip

\begin{center}
{\Large \bf {The symmetry and simplicity of the   \\
 }}
 \bigskip
{\Large \bf {laws of physics and the Higgs boson}}
    \bigskip
\bigskip
\end{center}

  \begin{center}

 \bf {   Juan Maldacena }
  \bigskip \rm
\bigskip
 
   Institute for Advanced Study,  Princeton, NJ 08540, USA  \\
\rm

\bigskip
\bigskip

  \end{center}

 \bigskip\bigskip
  \begin{abstract}

 We describe the theoretical ideas, developed  between the 1950s-1970s,
which led to the prediction of the Higgs boson, the particle that was discovered
in 2012. 
The forces of nature are based on
symmetry principles. We explain the nature of these symmetries through an economic analogy. 
We also discuss the  Higgs mechanism, which  is necessary to avoid some of
the naive consequences of these symmetries,  and to explain
various features of elementary particles.

 \medskip
  \noindent
  \end{abstract}
\bigskip \bigskip \bigskip
\begin{center}
\includegraphics[scale=.3]{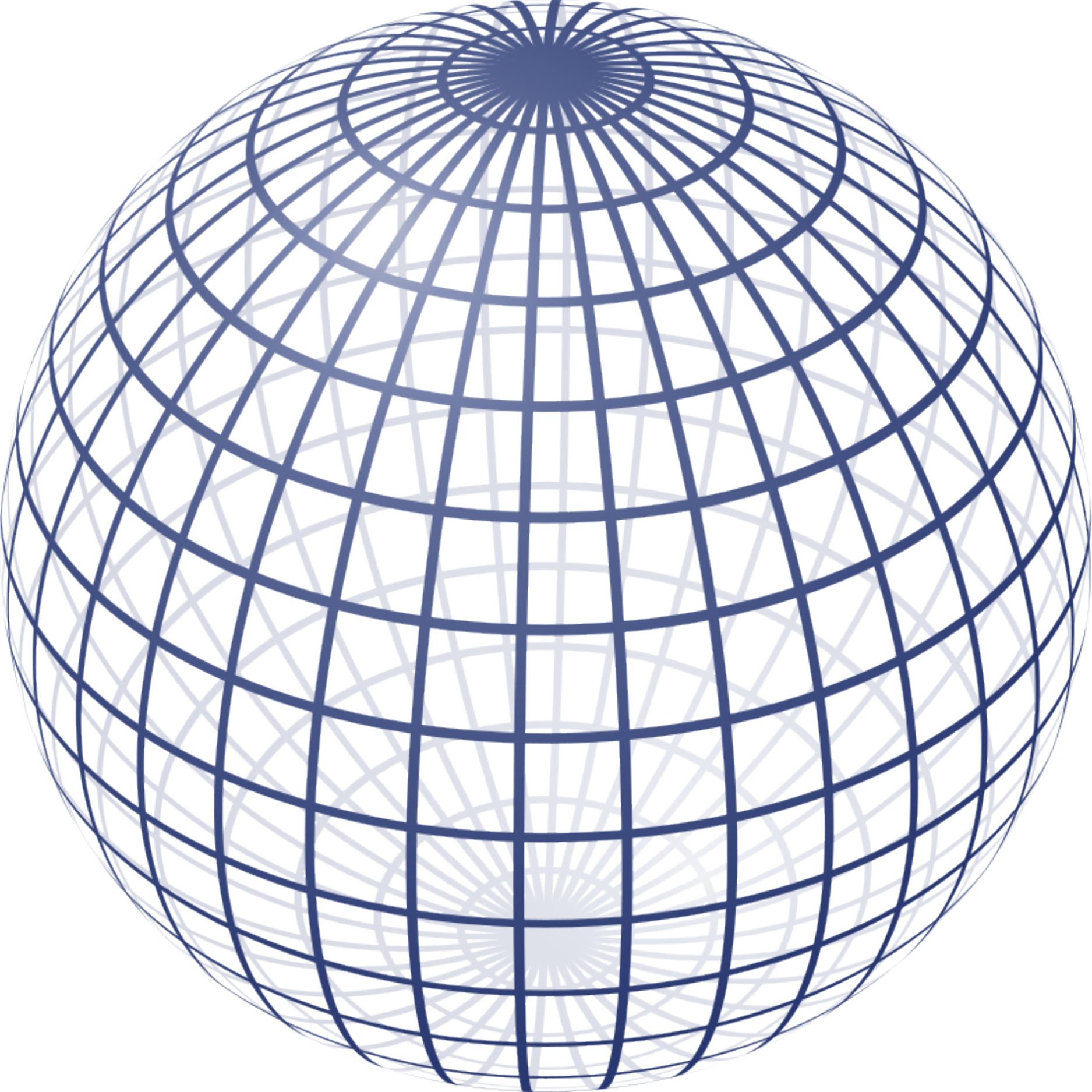}
\end{center}

  \end{titlepage}



\section{A Fairy Tale}

Our present understanding of particle physics is like the story of the Beauty and the Beast. 
Beauty represents the forces of nature: electromagnetism, the weak force, the strong force and 
gravity.  These are all based on a symmetry principle, called gauge symmetry. 
In addition,  we also need the   beast, which is the 
 so-called Higgs field.  It contains much of the mysterious and  strange (some would say ``ugly'') aspects  
  of particle physics.  
But we  need both  to describe nature because we are the children of this marriage between 
the beauty and the beast.

\begin{figure}[h]
\begin{center}
\includegraphics[scale=.6]{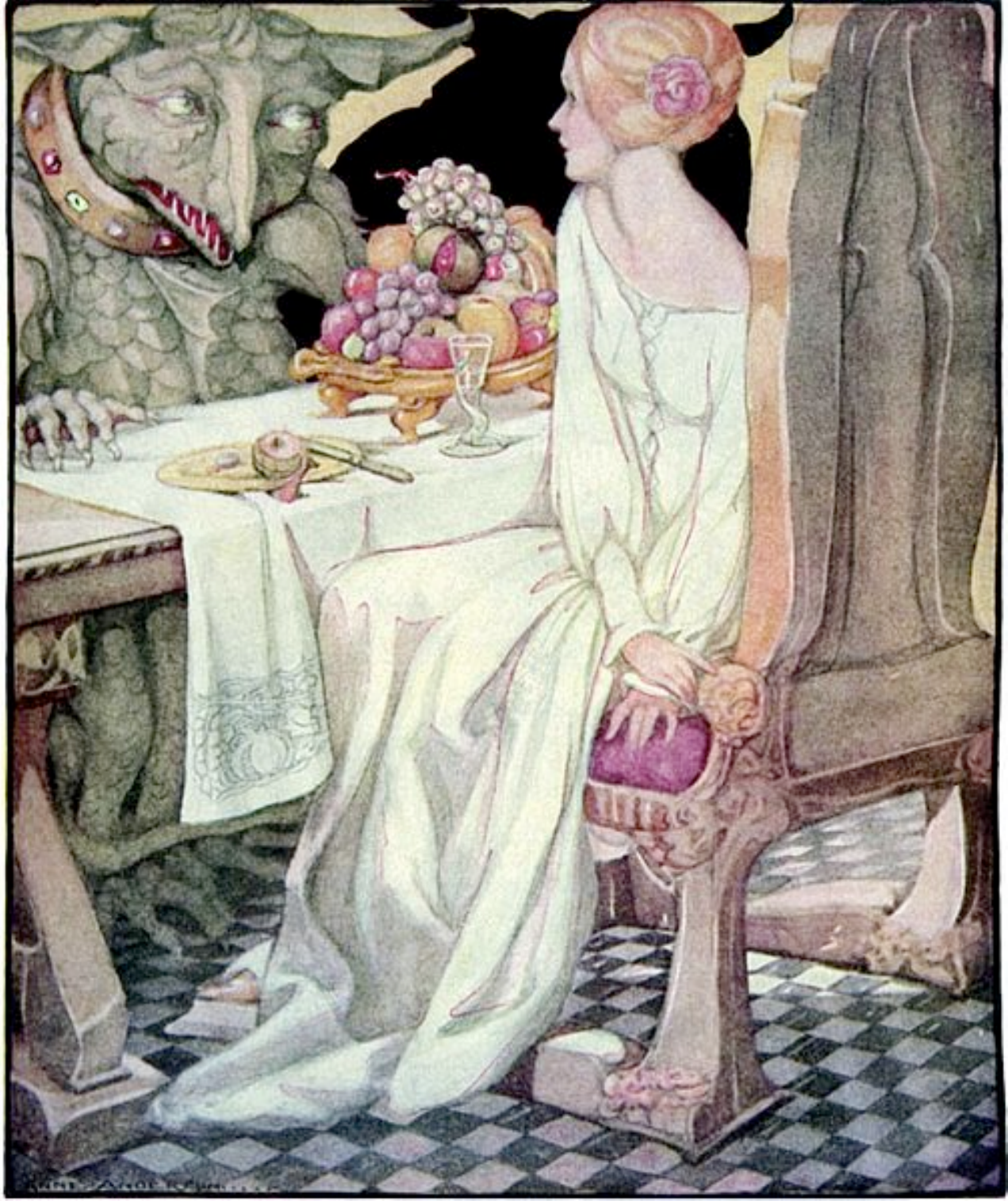}
\caption{  Particle physics is like the Beauty and the Beast. The four fundamental forces are Beauty and the Higgs field is
the Beast.  }
\label{BeautyBeast}
\end{center}
\end{figure}

Here we will attempt to introduce these two main actors: the beauty or gauge principle, and the beast or Higgs 
mechanism. We will do this through some analogies. 
  There have been many attempts to explain particle physics  to the general 
public. The usual expositions take the reader through a long path 
starting from the world of ordinary experience, through molecules, atoms, nucleus, etc, usually 
following the  order of historical discoveries. 
Here we will attempt something different, we will just parachute into the modern description. 
We will find ourselves in a sort of fairyland, with very simple rules. 
Our main point is to emphasize the role of gauge symmetries and to highlight the surprising simplicity of
these laws.  These are the laws of the ``Standard Model'' of particle physics.  They are  surprisingly 
simple given that they ultimately explain  most of the ordinary phenomena. 
They describe the universe since it was a  millisecond  old till today. They  also ultimately 
describe  almost all of known matter  (dark matter is the exception). 
 This description of nature is the result of arduous experimental work, which we will not review here, but
 can be found in many of the other popular descriptions of particle physics.

\section{Symmetry} 

\subsection{Ordinary symmetries}

\begin{figure}[h]
\begin{center}
\includegraphics[scale=.5]{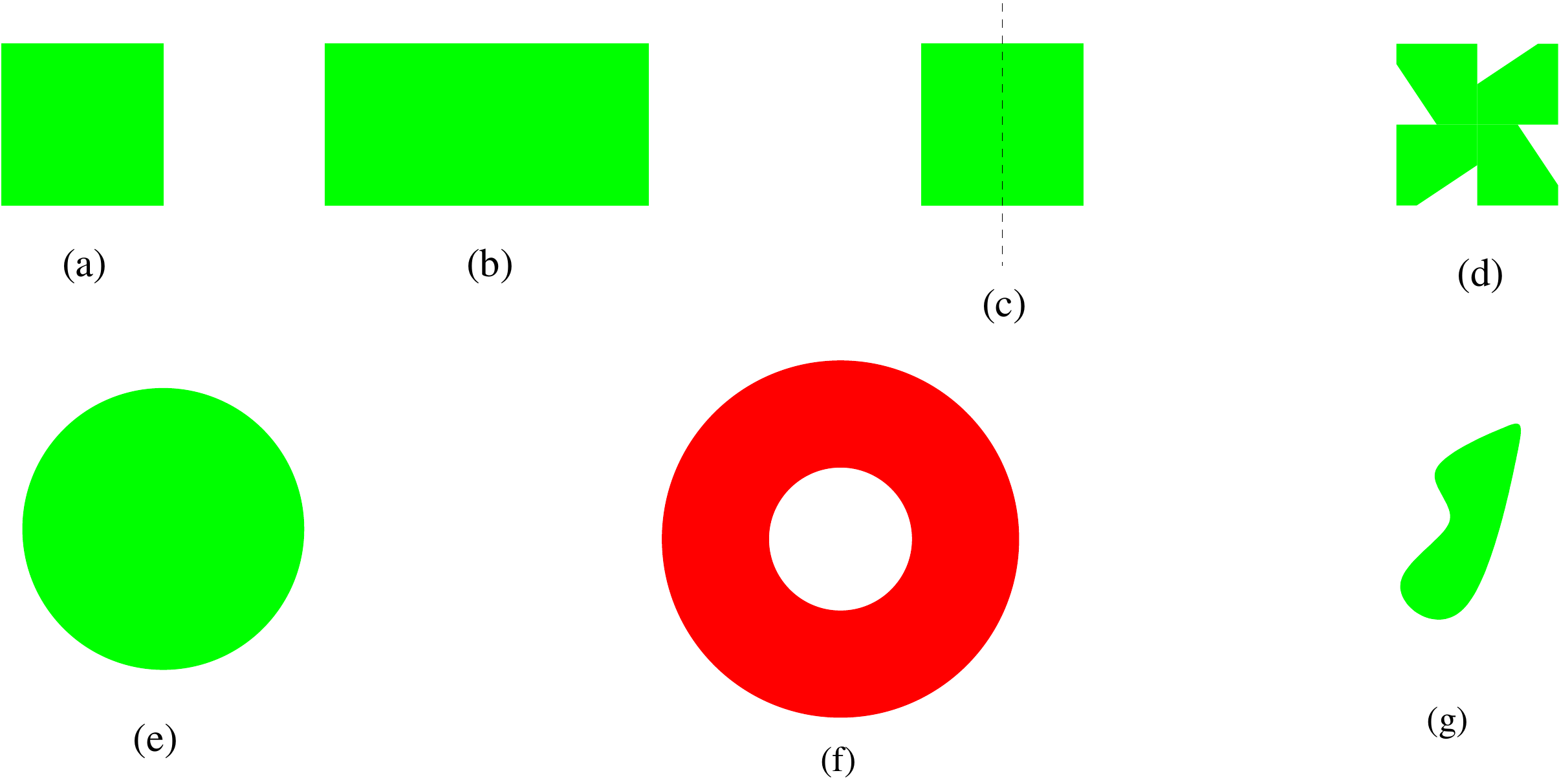}
\caption{Figures with various amounts of symmetry. (a) The square is symmetric under rotations
 by 90 degrees. (b) This rectangle is only symmetric under rotations by 180 degrees. (c) The square is also 
 symmetric under a reflection along the dotted line. (d) This figure is not symmetric under any reflection. However, it
is symmetric under rotations by 90 degrees. (e) This circle can be 
 rotated by any angle. (f) This ring has the same symmetries as the circle.   (g) This figure has no symmetry. }
\label{SymmetricFigures}
\end{center}
\end{figure}

First, 
 let us say a few words about  the notion of symmetry. Symmetry in physics is exactly the notion 
that we associate  with  symmetry in the colloquial language. It is a transformation that leaves an 
object unchanged.   For example, a square can be rotated 
by 90 degrees and it looks the same. If we had a rectangle, this would not be the case. 
See figure \ref{SymmetricFigures}. The rectangle can be rotated by 180 degrees. We can also 
have figures that have no symmetry, see figure \ref{SymmetricFigures}(c). 
A circle is very symmetric, it can be rotated by any angle and it remains the same.  Two different 
 figures can  have the 
same symmetries. For example, a circle and a ring are different figures but they have exactly the same 
symmetries. A situation could arise where we know what the symmetries of the object are, but we might 
not know what the object really is. This is sometimes enough to make predictions. For example, we
can have a rigid object which is symmetric under rotations, with the symmetries of a circle.
 Then we know that it will roll smoothly 
on a table. It can be a solid cylinder or a hollow cylinder, but both will roll smoothly on a table. 
Of course, in other respects the hollow and solid cylinder can behave differently. For example, one can float in 
water and the other might not. 

The symmetries we will talk about are a generalization of these more familiar ones. Their consequences will 
be to determine the forces of nature.

\subsection{  Electricity and magnetism reminder } 
 
\begin{figure}[h]
\begin{center}
\includegraphics[scale=.5]{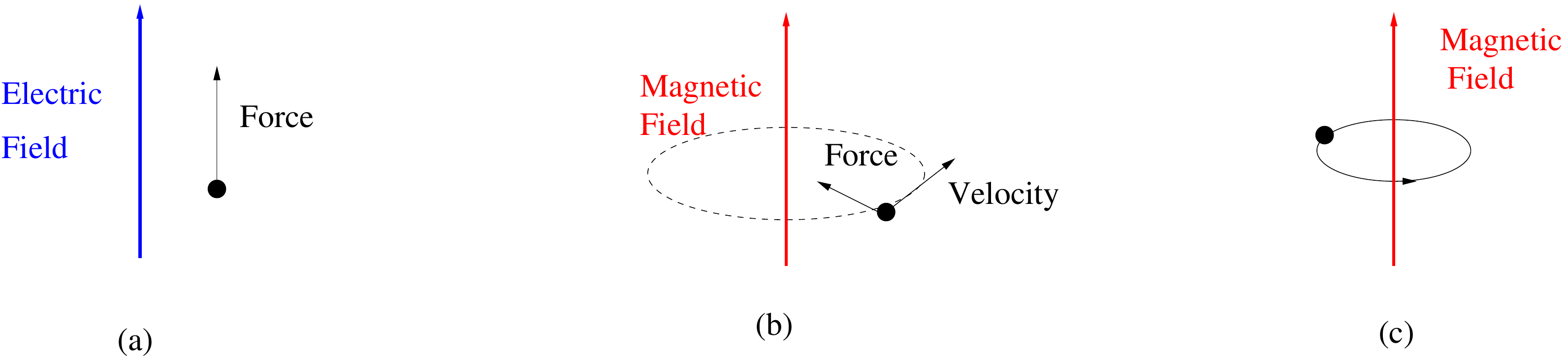}
\caption{  (a) An electric field exerts a force on a charged particle along the direction of the field.  (b)
 A magnetic field exerts a force on a moving charged particle that is perpendicular both to the magnetic field and the direction of motion. (c) 
 The  charged particle ends up moving in a circle around the magnetic field.     }
\label{ElectricMagnetic}
\end{center}
\end{figure}

Before starting our discussion, 
let us recall some facts about   electromagnetism.  
One postulates the existence of electric and magnetic fields. We think of them as little arrows 
at each point in spacetime. 
We feel the presence of these fields by their action on charged particles.   
 Electric fields act on 
charged particles by  pushing them along the direction of the electric field.  
 Magnetic fields only act on moving charges. In the presence of a magnetic field a moving charge 
feels a force that acts perpendicularly to the direction of its velocity. So if you were a charged particle 
and you are moving forwards in a vertical magnetic field, you would feel a force pushing you to your side. 
If there were no other forces, then you would end up moving along a circular trajectory. 
See figure \ref{ElectricMagnetic}. In summary, charged particles move in circles around magnetic fields. 

These electric and magnetic fields have their own dynamics. We can have electromagnetic waves, which 
are interlocked oscillations of electric and magnetic fields. These can propagate through the vacuum. 
They are radio waves, light, X-rays, gamma rays, etc. All this is   hopefully familiar to
you.  Do not panic!  You do not need to   know the detailed 
equations to understand what follows.
 All you need to know is that there are electric and magnetic fields, which can exist in otherwise 
empty space. These fields act on charged particles and affect their motion. Electric fields push them 
along the direction of the electric field. Charged particles move in circles around magnetic fields.

\subsection{Gauge symmetry}

Electromagnetism can be viewed as a ``gauge theory''. This is another point of view on 
electromagnetism. This point of view is particularly useful for generalizing it to other forces. It  is 
also useful for describing the quantum mechanical version. 
To explain what  a gauge symmetry is,
 it is convenient to introduce an economic analogy\footnote{ 
 The analogy between foreign exchange and lattice gauge theory was noted in  
K. Young, ``Foreign exchange market as a lattice gauge theory'', American Journal of Physics 67, 862 (1999). 
Here we will extend  that discussion, with the physics goal in mind. See also K. Illinski, http://arxiv.org/abs/hep-th/9705086 and P. Malaney, Harvard Ph.D. thesis. }.  In this economic analogy 
we will make a few idealizations and simplifications. 
Keep in mind that our goal is {\it not } to explain the real economy. Our
goal is to explain the real physical world. The good news is that the model is much simpler than the 
real economy. This is why physics is simpler than economics!
 
\begin{figure}[h]
\begin{center}
\includegraphics[scale=.5]{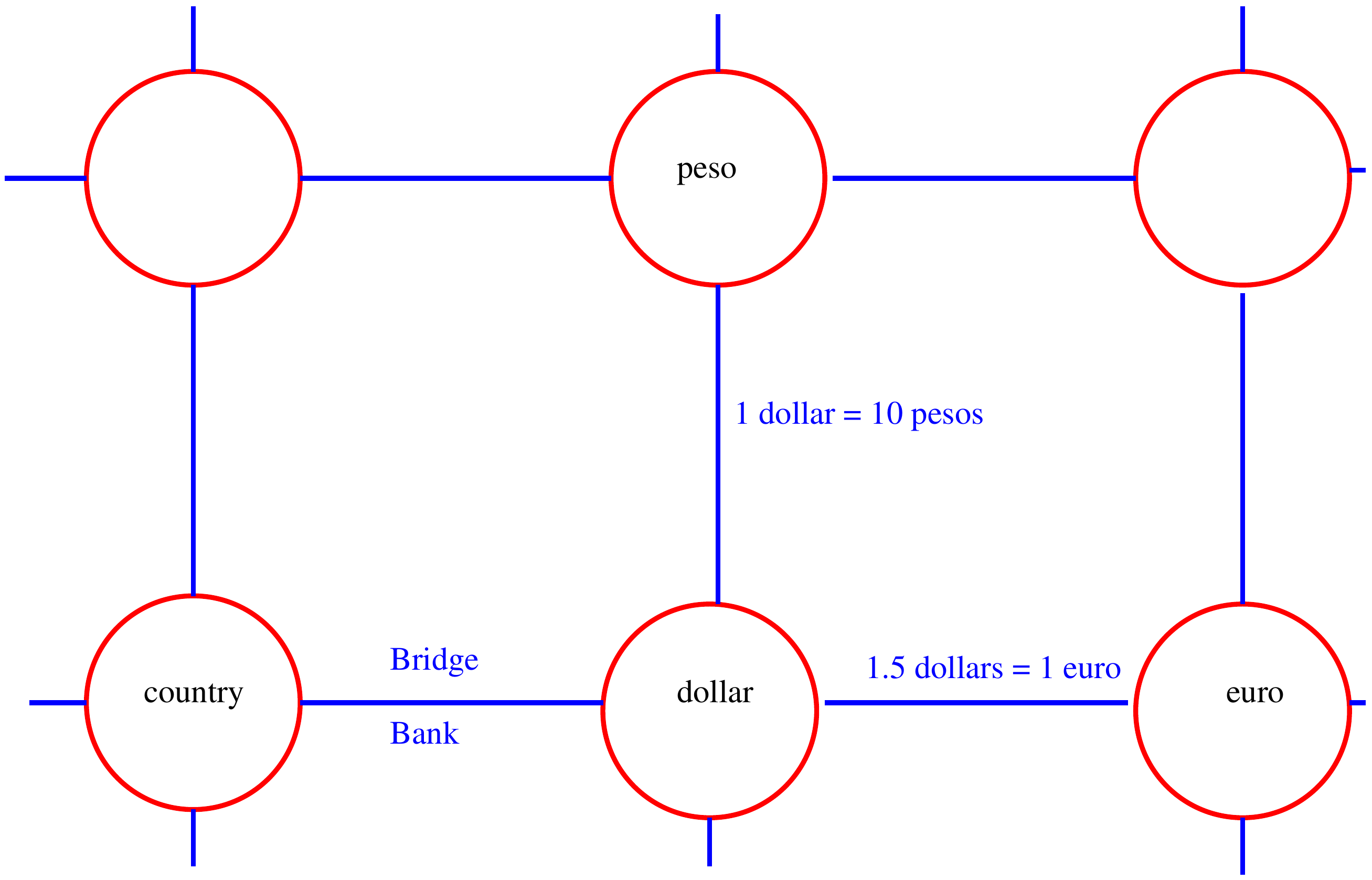}
\caption{  Each circle is a country that has its own currency. They are connected by blue bridges. At these bridges there is 
a bank. When you cross the bridge, you should change all your money to the new currency.   We have
indicated some of the currencies and  exchange rates. At each  bridge  there is an independent exchange rate.  }
\label{CountryGrid}
\end{center}
\end{figure}

Now, let us get to the  economic model. 
We imagine we have some countries. Each country has its own currency. Let us imagine 
that the countries are arranged on a regular grid on a flat world.   See figure \ref{CountryGrid}. 
Each country is connected with its neighbors with a bridge. 
At the bridge there is a bank. There you are required to change the money you are carrying into the 
new currency, the currency of the country you are crossing into. 
There is an independent bank at each bridge.  
  There is 
no central authority coordinating all the exchange rates between the various countries. Each bank 
is autonomous and sets the exchange rate  in an arbitrary way. 
 The bank charges no commission. 
For example, assume that  the currency in your original country is dollars and the one in the new country is 
euros. Suppose that the exchange rate posted by the bank at the bridge between two countries is 
  1.5 dollar = 1 euro. See figure \ref{CountryGrid}. Then if you have 
15 dollars the bank converts it to 10 euros as you cross the border. If you decide to come back 
your 10 euros will be converted to 15 dollars.  Therefore,  if you go to a neighboring country and 
you come right back, you end up with your original amount of money. Another rule is that 
you can only go from one country to the neighboring country. From there you can continue to any 
of its neighbors and so on. However, you cannot fly from one country to a distant country 
without passing through the intermediate ones. You can only walk from one to the next, crossing the 
various bridges and changing your money to the various currencies of the intermediate countries. 
The final assumption  is  that the only thing you can carry from one country to the next is money. 
You cannot carry gold, silver, or any other good.  We will later relax this 
assumption, but for now we will consider this simplest situation. 

Let us review the assumptions again. We have countries arranged in a grid. Each country 
has its own currency. The countries are connected through bridges. There is a bank at each bridge 
which exchanges money. The banks choose any exchange rate that they please. 
There is no commission. You can only carry money from one country to the next.  
  This  is fairly simple. All you can do is travel between countries 
converting your currency each time you cross a border.
\begin{figure}[h]
\begin{center}
\includegraphics[scale=.43]{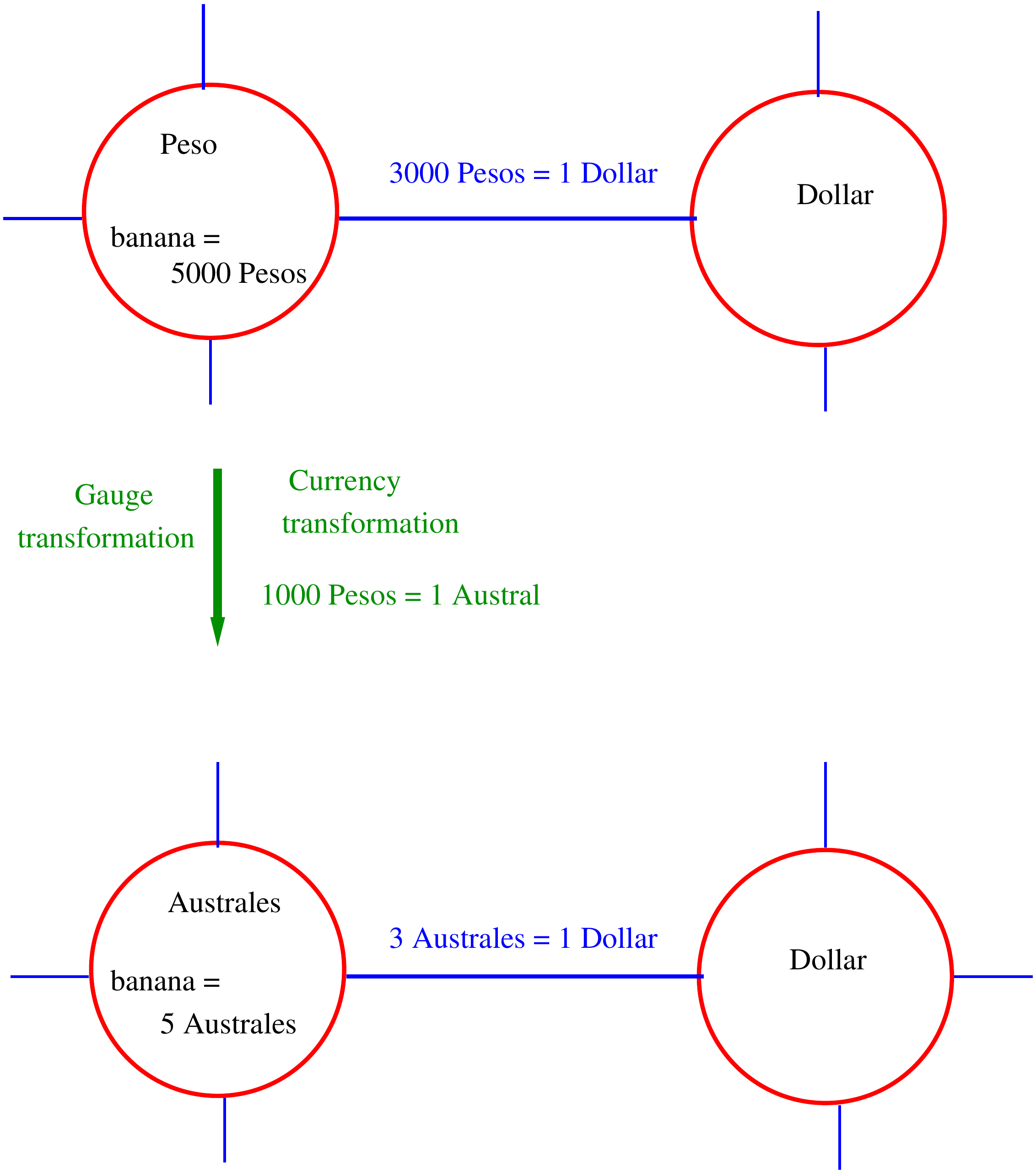}
\caption{ Each country can change its currency units. Here the country that was using Pesos changes the currency to 
Australes so that 1000 Pesos = 1 Austral. All prices and exchange rates change accordingly. Here we indicated how the price
of a banana would change. We have also indicated how the exchange rate to one of the neighboring countries changes. Of course, the exchange
rates with all  the  neighbors change. }
\label{PesosAustrales}
\end{center}
\end{figure}

\begin{figure}[h]
\begin{center}
\includegraphics[scale=.60]{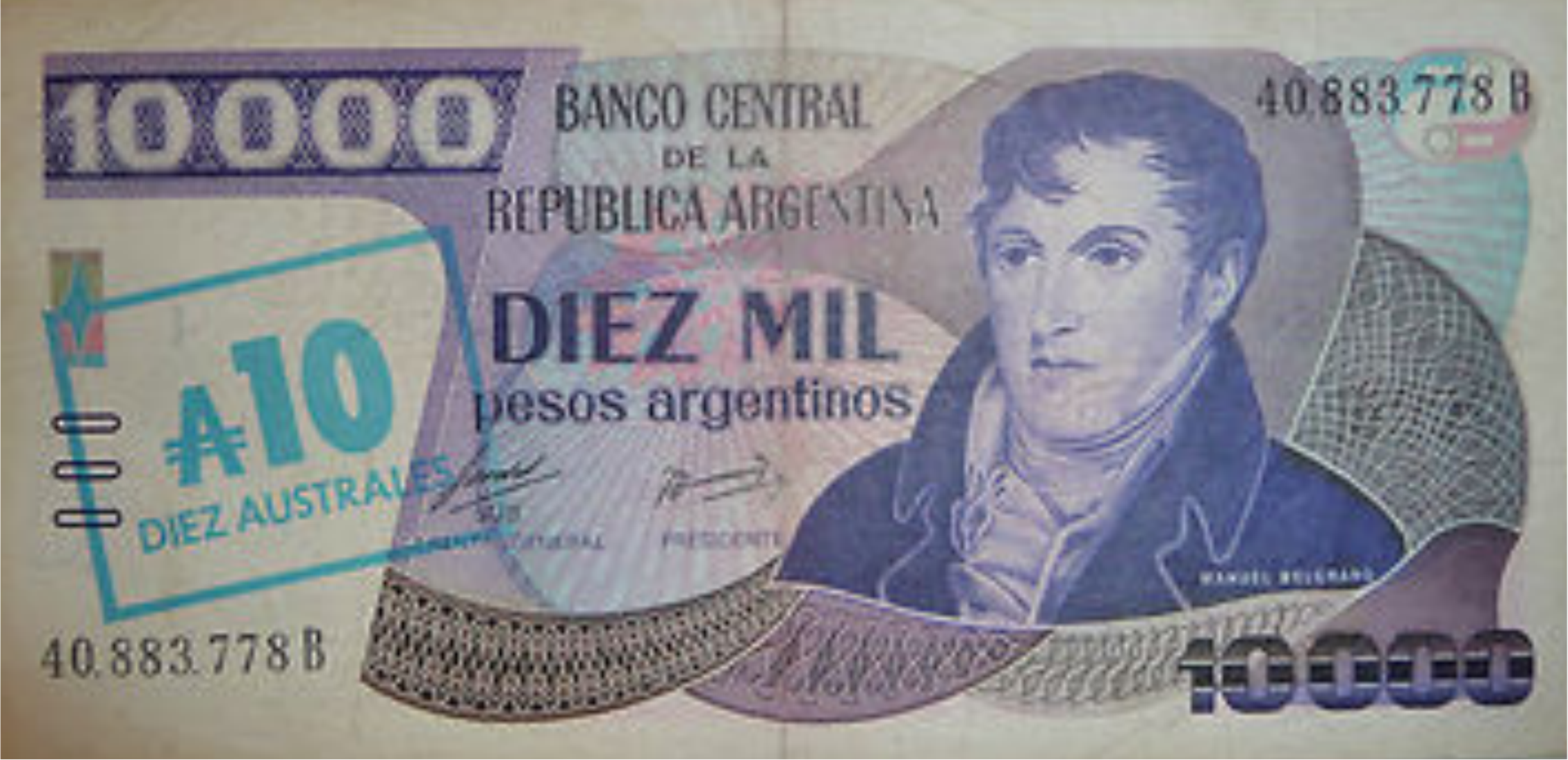}
\caption{  A gauge symmetry in action. Here we see a real world  change from Pesos Argentinos to Australes.   }
\label{Billete}
\end{center}
\end{figure}

Where is the symmetry? The gauge symmetry is the following. Imagine that one of the countries 
has accumulated too many zeros in its currency and wants to drop them. This is fairly common in 
the real world in countries with high inflation. What happens is that one day the local government
 decides that
they will change their currency units. For example, instead of using  Pesos now everybody needs to use  ``Australes''. 
The government declares  1,000 Pesos will now be worth 1 Austral, or 1,000 Pesos = 1 Austral.
So everybody changes all prices and exchange rates accordingly. If  you needed to pay 5,000 Pesos  for a banana, now you
will  need to pay 5 Australes. If your salary was 1 million Pesos, it will now be 1 thousand Australes. 
Suppose the neighboring country is the USA. If the exchange rate was 3,000 Pesos = 1 Dollar,  
it will now be 3 Australes = 1 Dollar.    See figure \ref{PesosAustrales}. 
We call this a ``symmetry''  because after this change nothing really changes, nobody is richer or 
poorer and the change offers no new economic opportunities. It is   done purely for convenience. 
You can see 
this gauge symmetry in action in some Argentinean banknotes in figure \ref{Billete}. 
 It is called a ``gauge'' symmetry because it is  a symmetry of the units we use to measure or ``gauge''
the value of various quantities. 
 
This symmetry is  ``local'', in the sense that each country 
can locally decide to perform this change, independently of what the neighboring countries decide to do. 
Some countries might like to do it more frequently than others. In the real world,  Argentina 
has eliminated thirteen zeros through various actions of this   ``gauge symmetry"  since the 1960s, so 
that 1 Peso of today = $10^{13}$ Pesos of the 1960s.

Now we will focus on speculators. A speculator is somebody who will travel along various routes 
converting his money as he crosses each bridge. His goal is to earn money. He wants to  travel along 
the    paths with the highest monetary gain. Recall that according to our rules he has to actually travel 
to the different countries. He cannot sit at a desk and order his trades on a computer. 

Do you think you can make money in this world? 
Think about it, what would you look for?

At first sight it seems that you cannot make any  money. In fact,  if you 
go from one country to its neighbor and back,  you end up with the same amount of money.   
However, it is possible to make money if you come back a different way!
 
 \begin{figure}[h]
\begin{center}
\includegraphics[scale=.60]{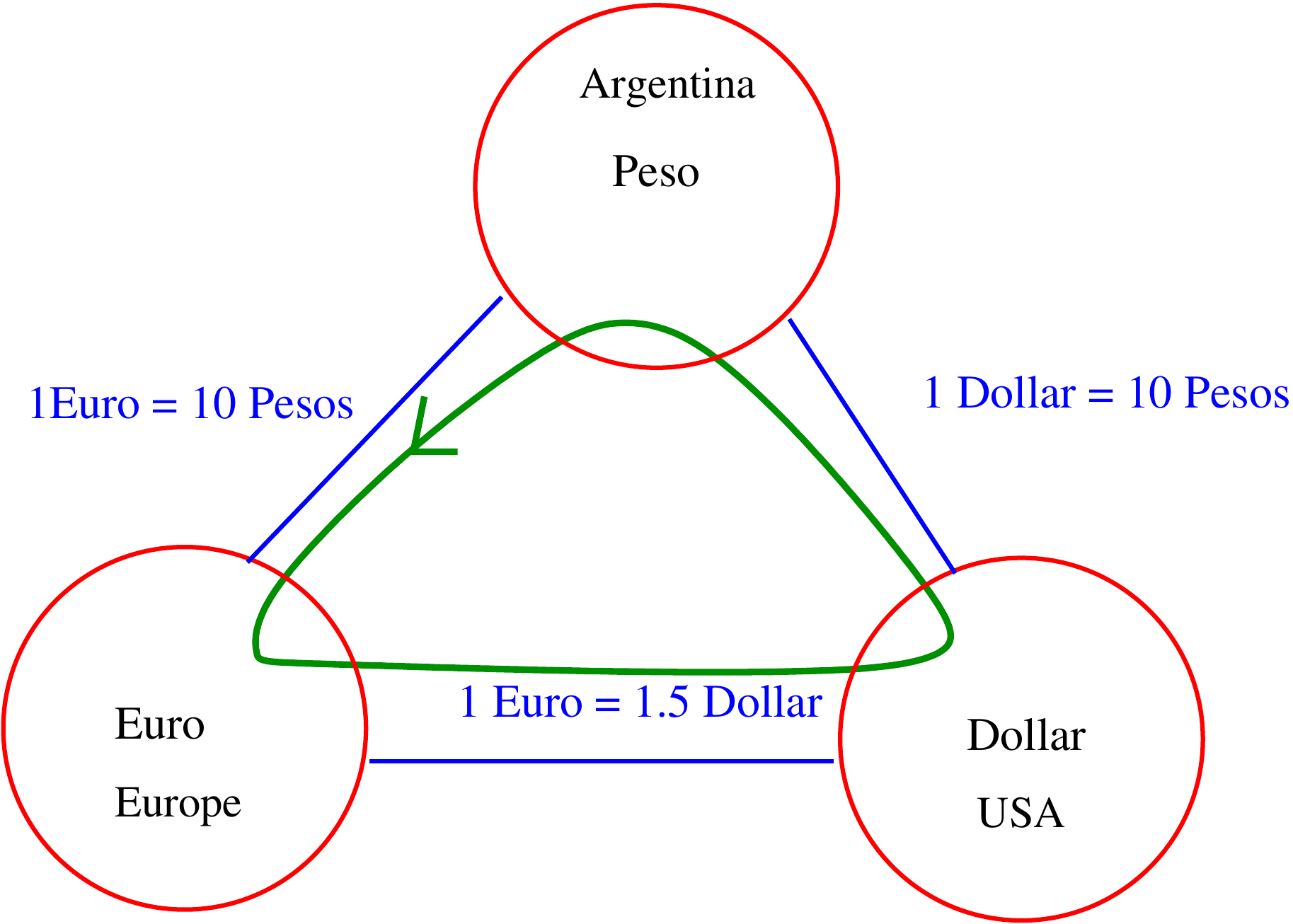}
\caption{  Here we see three countries with the respective currencies. On the blue bridges we see the corresponding 
exchange rates. If you follow the circuit along the green line, you would earn money. You would have a gain factor of 1.5 or 
earnings of 50\% .    }
\label{ThreeCountries}
\end{center}
\end{figure}

 Just as an example let us 
consider three countries, say the USA,   Europe and Argentina, with their corresponding currencies
dollars, euros, and pesos. Now let us imagine that there are bridges connecting these three countries, 
see figure \ref{ThreeCountries}. 
The exchange rates are as follows:
\be \notag
 1.5  \, { \rm dollars} = 1 \,  {\rm euro} ~,~~~~~~~ 1 \, {\rm  dollar}  = 10 \, {\rm pesos}~,~~~~~~~~~~
   1 \, {\rm euro} = 10 \, {\rm  pesos}  
  \ee 
In a situation like this, can you make money? Think about it before you continue reading. 
It is worth the effort!

We can earn money as follows. We   start in Argentina with 10 pesos.  We go to Europe and get 1 euro.  We go to the USA and get 
1.5 dollars. And then   back to Argentina and we get 15 pesos. We started with ten and ended with fifteen. 
The gain of this circuit  is a factor of 1.5, or earnings of 50\%.  If you started the circuit with $X$ pesos, you 
would have $1.5 \times X $ pesos at the end. This factor is independent of the currency units. If
the Argentinean government changes from Pesos to Australes, then the gain of the circuit is still 
the same, it is still a factor of 1.5. 

You might think that the banks were dumb  to  set the ``wrong'' exchange rates and that speculators
are taking advantage of them. Well, this is according to the rules we have enunciated. The banks 
set the exchange rates they want. With some choices there will be opportunities to speculate and with 
some others there will not be. 
The speculators have a very simple mind, they only care about earning money and they will choose
the path that makes them earn most money. 
In the above 
 situation the speculators would be moving in circles going from Argentina to Europe to the USA and back 
to Argentina.  They would be following the green line in figure \ref{ThreeCountries}.

Now, in physics the countries are analogous to points, or small regions, in space. 
The whole set of exchange rates is a configuration of the magnetic potentials throughout space. 
A situation  
like the one in figure \ref{ThreeCountries}, where you can earn money, is called a magnetic field. The amount of gain is 
related to the magnetic field. 
 The speculators are called electrons or charged particles. In the presence of magnetic 
fields, they simply move in circles in order to earn money. 
In fact, the total gain along the circuit is the flux of the magnetic field through the area enclosed
 by the circle.  
 Now imagine that you are a speculator that has  debt  instead of having money. In that 
case you would go around these countries in the opposite direction! Then your debts would be reduced
in the same proportion. In the example of figure \ref{ThreeCountries}, 
 your debts would be reduced by 
a factor of $1/1.5$ by circulating in the direction opposite to the green arrow. 
 In physics, we have positrons, which are particles like the electron 
but with the opposite charge. In fact, in a magnetic field 
positrons  circulate in the opposite direction as compared to electrons.

\begin{figure}[h]
\begin{center}
\includegraphics[scale=.35]{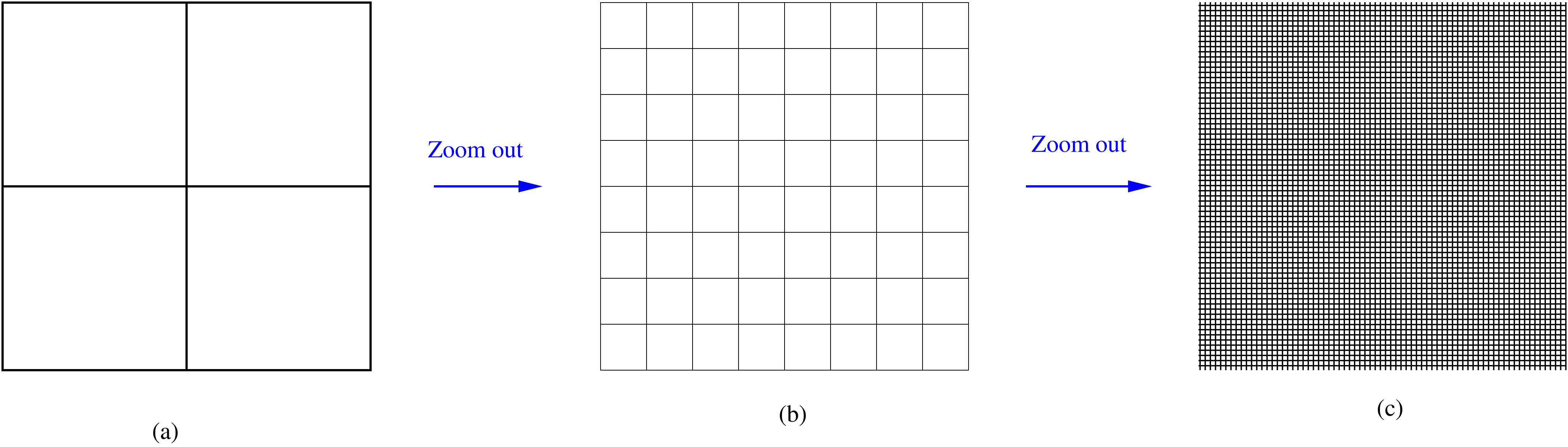}
\caption{  Here we display the grid of countries at various scales. We are zooming out as we move to the right. 
If we zoom out sufficiently we can view the whole grid as a continuum.  }
\label{Continuum}
\end{center}
\end{figure}

In physics, we imagine that this
 story about countries and exchange rates is happening at very, very short distances, much 
shorter than the ones we can measure today. When we look at any physical system, even empty space, we are looking at 
all these countries from very far away, so that they look like a continuum. 
 See figure \ref{Continuum}. When an electron is moving in the vacuum,   it  is 
seamlessly moving from a point in spacetime to the next. In the very microscopic description, it would be constantly changing
between the different countries, changing the money it is carrying, and becoming ``richer'' in the process. 
In physics we do not know whether there is an underlying discrete structure like the countries we have described. However, 
when we do computations in gauge theories we often assume a discrete structure like this one and then take the continuum 
limit when all the 
countries are very close to each other.

\begin{figure}[h]
\begin{center}
\includegraphics[scale=.38]{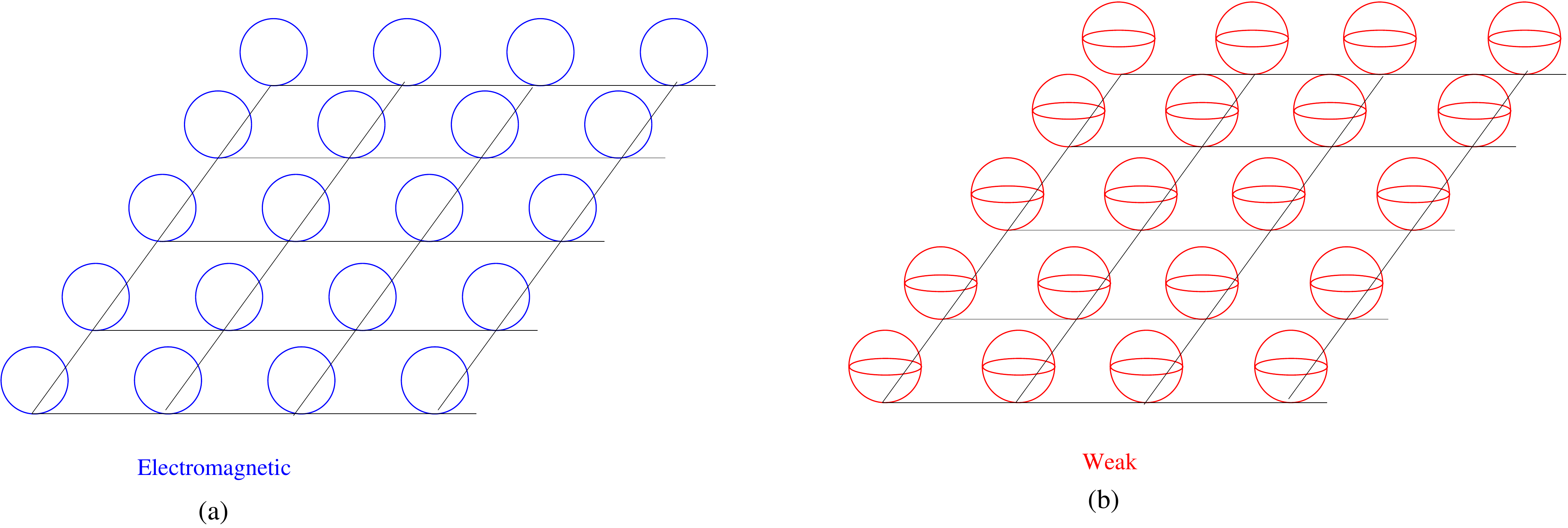}
\caption{ (a) The electromagnetic interaction 
 has the same symmetries as a configuration where we have a circle at each point
of spacetime. Here each spacetime point is where the black  lines intersect. We can think of the circle as an extra dimension.
 (b) The weak force has the same 
symmetries as a configuration 
 where we have a sphere at each point in spacetime. We do not know whether the circles or spheres really exist as extra dimensions. 
What we do know is that the gauge symmetry is the same as if they existed.
 The circles of spheres are useful for visualization but we only think about their symmetry and focus only on the  associated 
 ``exchange rates''. 
   }
\label{CirclesSpheres}
\end{center}
\end{figure}

Electromagnetism is based on a similar gauge symmetry. In fact, at each point in spacetime the symmetry corresponds to 
the symmetry of rotations of a circle. One way to picture it is to imagine that at each point in spacetime we have an extra circle, an 
extra dimension. See figure \ref{CirclesSpheres}(a). 
 The ``country'' that is located at each point in spacetime chooses a way to define angles on this extra circle in an 
independent way. More precisely, 
each ``country'' chooses a point on the circle that they call ``zero angle'' and then describe the position of any 
other point in terms of the angle relative to this point. 
This is like choosing the currency in the  economic  example. 
Now, in physics, we do not know whether  this circle is real. We do not know if indeed there is an extra dimension. All we know is that
the symmetry is similar to the symmetry we would have if there was an extra dimension.
 In physics we like to make as few assumptions as possible. 
An extra dimension is not a necessary assumption, only the symmetry is. Also the only relevant quantities are the magnetic potentials which 
tell us how the position of a particle in the extra circle changes as we go from one point in spacetime to its neighbor. 
 
 In electromagnetism the electric and magnetic fields obey some equations, the so-called Maxwell equations. 
 In the economic analogy this would be analogous to a requirement on the exchange rates. In the economic model we can 
 intuitively understand   this requirement as follows. Let us imagine we have a configuration with generic 
exchange rates. Speculators start carrying money around. 
  Suppose 
 we focus on a particular bridge, where   a particular 
 bank sits. There will be speculators crossing this bridge in both directions. However, if there are more speculators going in one direction than
 in the other direction, then the bank might run out of one of the currencies. 
 For example,  consider the bank sitting  at a bridge that connects Pesos to dollars. 
 If there are more speculators wanting to buy dollars than there are speculators wanting to buy pesos,  the bank will run out of dollars. 
 If this happened in the real world, then the bank would adjust the exchange rate so that there would be
 fewer speculators wanting to buy dollars. 
 In fact, if we assume that the number of speculators following a particular circuit is proportional to the gain that they will have along this circle, then
 one finds that the condition for banks   not to  run out of either of the currencies, or that the net flow of money across each bridge is zero, is 
 equivalent to Maxwell's equations. 
 The  mathematically inclined  reader can find the derivation in the appendix.  

\subsection{The weak force } 

Let us now turn to the weak force. 
  It is the force 
 responsible for radioactive 
decays. For example,  a free neutron (outside a nucleus)  decays in about 15 minutes   to 
a proton, an electron and a neutrino. This  is a  very slow decay compared to other 
processes that happen on microscopic time scales. The weak force is not terribly relevant for our
everyday life. However, despite its weakness, it played an important role in the history of the universe. More
specifically, in the synthesis of the chemical elements in stars. In fact, all the chemical elements around us, except for hydrogen and helium, were
``cooked'' in stars.  The weak force played a crucial role in this process.  
 Closer to home, we can say that the weak force can move mountains!
More precisely, weak decays inside the earth are partly responsible for maintaining the earth 
hot, which in turn moves the continents,  creating  the mountains!

The weak force can also be understood using a gauge theory. In this case, at each point in space 
we have the symmetries of a sphere, let us call it the weak sphere. See figure \ref{CirclesSpheres}(b). 
We do not know whether the 
sphere is real or not.    What we do know is that when we go from one point in space to another we have
to specify three quantities, three ``exchange rates''. Instead of carrying money, we are now carrying an object that has some
orientation in the weak sphere. If we start at a country with an object in the weak sphere,   as we go to the neighboring country we have 
to re-orient the object according to the ``weak exchange'' rates.    We need three quantities since we have to 
specify a rotation axis (two quantities) and an angle of rotation around that axis (the third quantity). 
So,   instead of one magnetic field, we have three different types of magnetic fields. 
There are equations, similar to the ones for electromagnetism, which govern the 
behavior of these magnetic fields together with  the corresponding electric ones. 
 These equations were first proposed by Yang and Mills in 1954. 
When  W. Pauli heard about it, he  strongly objected. Pauli said that the Yang-Mills
theory implied that there would be new massless particles,  which are not observed in nature. 
This was a beautiful theory killed by an ugly fact.

\begin{figure}[h]
\begin{center}
\includegraphics[scale=.6]{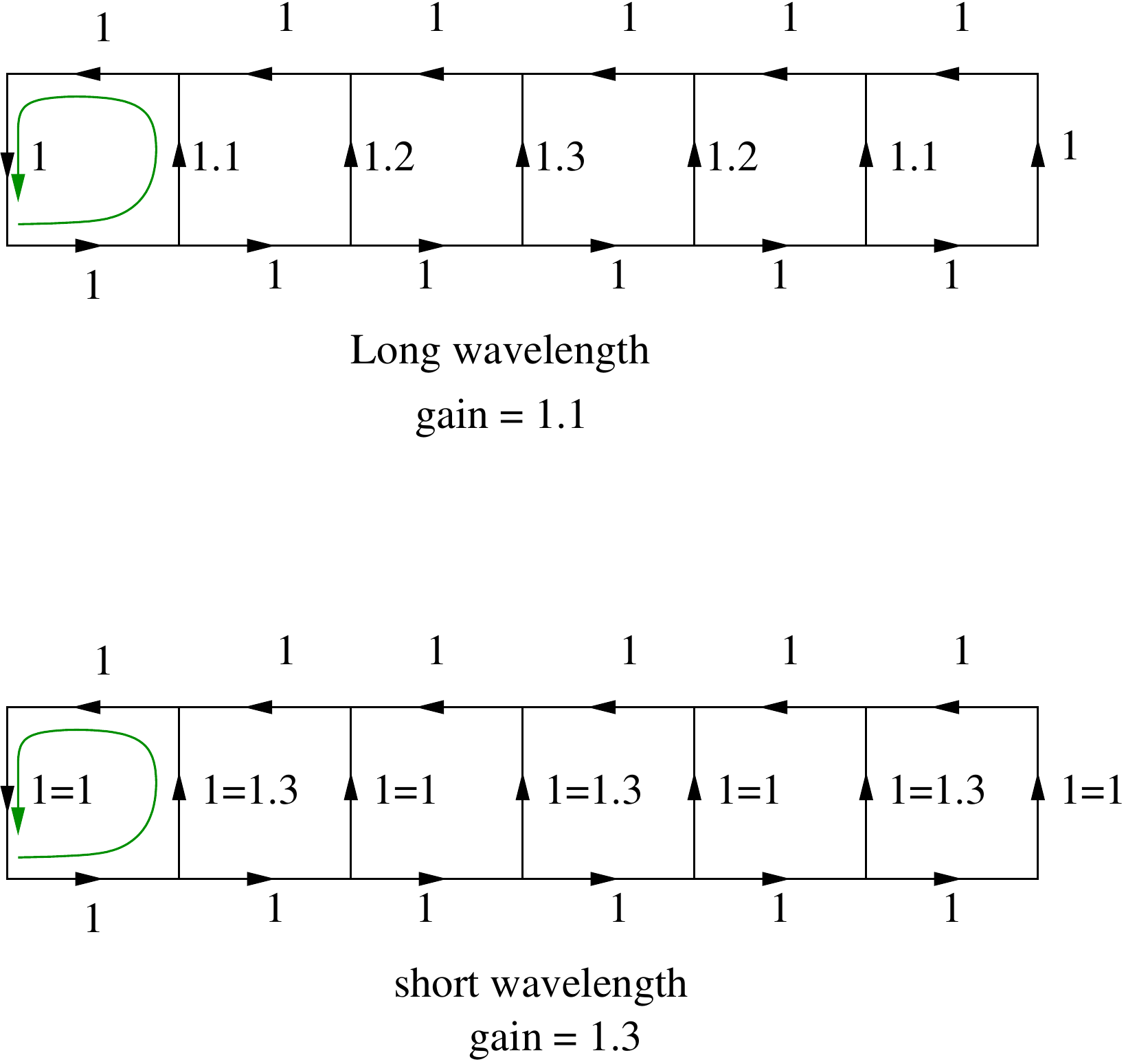}
\caption{  We can see a configuration of exchange rates with a long wavelength or a short wavelength. 
The wave consists in the fact that the numbers in the middle go up and down as we move from left to right. 
Each segment is a bank sitting between two countries. The countries sit at the line intersections. The number indicates the
exchange rate when you cross the bridge in the direction of the arrow.  Notice that the total amplitude of the 
wave is the same; the exchange rates go between 1 and 1.3.   The gain obtained by following an elementary square circle, indicated by the
green lines, is smaller for the longer wavelength configuration.  }
\label{LongShortWavelength}
\end{center}
\end{figure}

\subsection{Why massless particles? }
 
To understand Pauli's objection, let  us first focus on some properties of waves. 
In the systems we consider we can 
have waves with different wavelengths. The wavelength of a wave is the distance between two
successive maxima. 
In physical systems one is often interested in the energy cost for exciting waves of various 
wavelengths. For a wave with a  given amplitude this energy cost can depend on the wavelength. 
For an electromagnetic wave, this energy cost decreases when we make the wave longer and longer. 
 It turns out that 
the mass of the particle is related to the energy cost to excite a very long wavelength wave. 
This is related to the famous formula $E = mc^2$. Unfortunately I have not found a short way to explain this, so 
you will have to trust me on this. 
 In our economic analogy we have not talked about energy. Let us simply say that the energy increases as the gain 
 available to speculators increases.  This makes intuitive sense, the more the speculators can earn, the harder it is for the banks!
 Therefore configurations with less gain have a lower energy cost. 
 In figure \ref{LongShortWavelength} we describe a whole sequence of exchange rates 
with a long and a short wavelength configuration. The crucial point is that the gain that speculators obtain by following the 
elementary square circuits (denoted by the green arrow in figure \ref{LongShortWavelength}) is
only related to the difference between neighboring exchange rates, but not on their absolute magnitude. 
Therefore, the  longer the  wavelength, the smaller the difference. 
The fact that the gain becomes smaller as the wavelength becomes longer implies that the associated particle, the 
photon, is massless. This is a correct argument for electromagnetism and it is also  correct for the weak force for
the same basic reason.  At least it is true for the version of the weak force described so far...

 \section{The Higgs mechanism} 
 
\begin{figure}[h]
\begin{center}
\includegraphics[scale=.48]{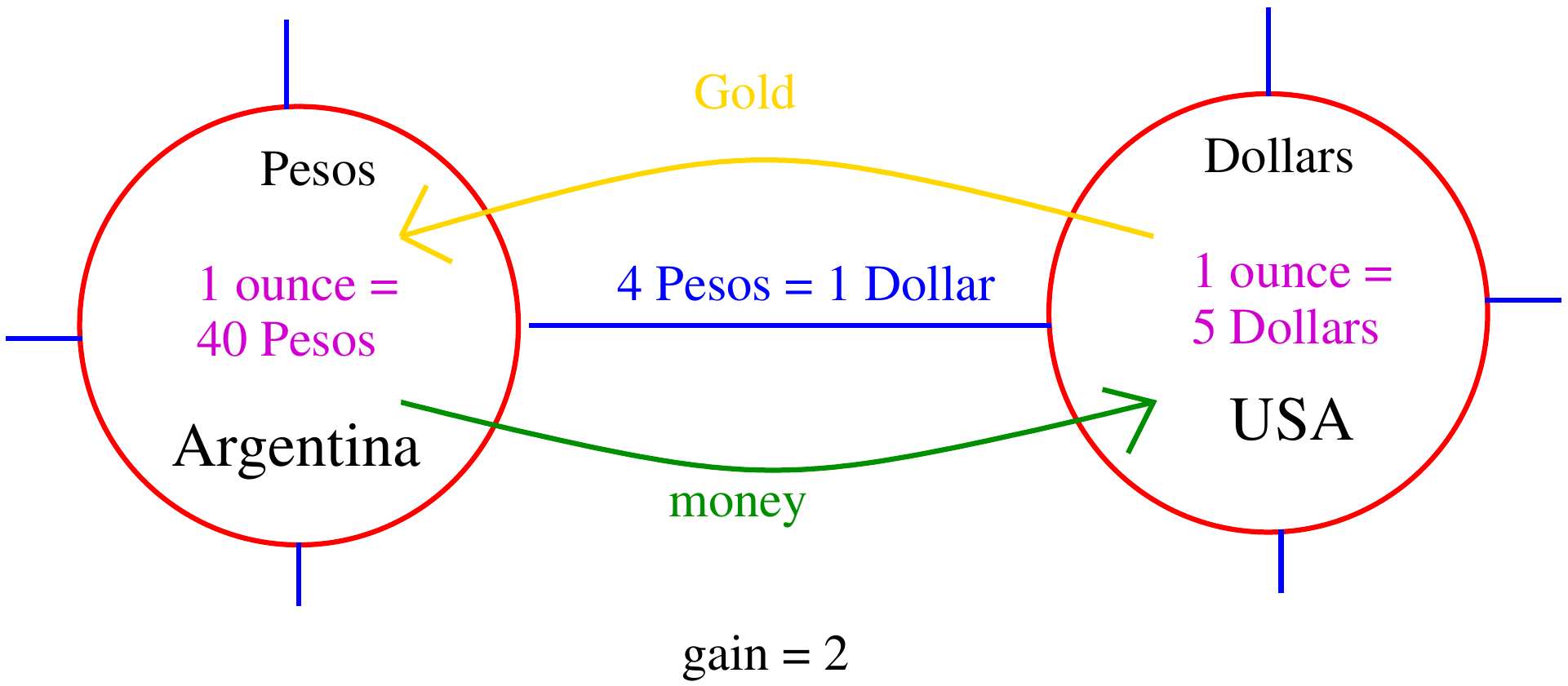}
\caption{   Here we see two countries with their respective gold prices. We have also given the exchange rate between them. 
For these prices and exchange rate it is possible to earn money by buying gold in the USA, taking it to Argentina, selling it there and 
bringing back the money, exchanging it at the bank, of course. The net gain is a factor of two or 100\%. Gold is indicated in yellow and money in 
green. }
\label{GoldExample}
\end{center}
\end{figure}

Now, it turns out that there is a way around  this argument. This is called the Higgs mechanism, which was 
proposed by many researchers\footnote{ 
These  include Anderson, Brout, Englert, Goldstone, Guralnik, Hagen,  Higgs, Kibble, Nambu, etc.  The detailed history can be found elsewhere.}. 
Here we will explain it using the economic analogy. So far, we have assumed that  you can only carry money 
between countries.  Now, let us assume that you are also allowed to carry gold. 
So the new rule is that you are allowed to carry gold and/or money between different countries. 
  Gold has a price in each country, which   is set by the inhabitants of each country independently 
  of the others.  
  A savvy speculator   realizes that   a new opportunity opens up. 
 You  can now buy gold in one country, take it to the next, sell it, and bring back the money to the first country. 
 As an example, say that the exchange rate between pesos and dollars is  4 pesos = 1 dollar. And the price of 
 gold in Argentina is 40 pesos per ounce and the price in the USA is  5 dollars per ounce. What would you do? 
 Again,  the prices and exchange rates are
 \be \notag
 4 \, {\rm pesos} = 1 \, {\rm dollar} ~,~~~~~~~~1 \, {\rm ounce} = 40 \, {\rm pesos} ~,~~~~~~~~1 \, {\rm ounce} = 5 \, {\rm dollars} 
 \ee
 Think about it, do not continue reading until you have the answer. It is a bit hard, but worth the effort!
 Yes, indeed, you would start with five dollars in the USA, buy gold there, go to Argentina, sell it for 40 pesos, 
 go back to the USA and get 10 dollars when you cross the bridge back.
  This operation then has a gain of a factor of two, or a 
 100\% profit. See figure \ref{GoldExample}. 
 Note that we continue to have the gauge symmetry. If the Argentinean government changes the currency to Australes, 
 your gain would be the same.  
 Now, we can use this gauge symmetry to choose the currency so that the price of gold is the same in all countries. 
 Let us call the new units new pesos and new dollars. Now the price of gold is 1 New peso per ounce and 1 New dollar per
 ounce. However, the exchange rates might not be one to one. In fact, they cannot be one to one if
originally there was an opportunity to speculate. For the example in figure \ref{GoldExample}  the new exchange rate is 
1 New peso = 2 New dollars. Note that this is {\it not} a ``gold standard''  that removes all exchange rates. 
It is very important that the exchange rates are still present. 

In summary, now the new prices and exchange rates are 
 \be \notag
 1 \, {\rm new ~peso} = 2 \, {\rm new~ dollars} ~,~~~~~~~~1 \, {\rm ounce} = 1 \, {\rm new~peso} ~,~~~~~~~~1 \, {\rm ounce} = 1 \, {\rm new ~dollars} 
 \ee
If you are a speculator, now it is easier to see what to do, isn't it?
With these new currency units obtained by setting the price of gold to one, one can immediately see that if any exchange rate is different than 
one to one, then there is an opportunity to speculate by performing the gold circuit.  
The opportunity to speculate remains, and the gain remains a factor of two, or 
100\% profit. As usual, the net gain does not change when we change currency units.

Now, a crucial new feature of this new economic model is that the gain does not become smaller and smaller
as we increase the wavelength. The reason is that now you check whether you can gain money by looking at a single 
bridge, and you do not need to compare neighboring bridges. Once we set the price of gold to one everywhere, then any 
exchange rate that is different from one to one leads to an opportunity to speculate. 
In the physics version, this means that it now costs some energy to move any exchange rate away from one to one. This cost is present even for long
wavelength configurations. In physics, this   leads to a massive particle,  to a massive photon. A mechanism of this 
sort is happening inside superconductors, and this was indeed an inspiration for the Higgs mechanism.

In physics, we think that  a similar mechanism gives 
  rise to a mass to the particles that mediate
the weak force. Instead of gold, we now have some object at each country, or spacetime point, 
that has some orientation in the weak sphere.
This is the so-called Higgs field. 
 We can use this object to completely fix the orientation of each weak sphere.  Now the object is oriented in the same way everywhere. 
 This is analogous to fixing the price of gold to one.  
  We still have the weak ``exchange rates'' which tell us how we get rotated in the weak sphere when we go from one country to the next. 
  The zero gain configuration is  when we are not rotated at all. In that case we can say that 
  the weak exchange rates are all ``one to one''. If we are rotated in any 
  way, there will be a possibility to speculate.  
 As a consequence the weak force mediators are massive. 
 In this way,  the heroic  Higgs field  has rescued the beautiful lady. We can explain the weak force using  gauge symmetry
and, at the same time, avoid having massless particles.

\subsection{Apologizing for one oversimplification} 

We have made an implicit oversimplification in this description, which can 
  confuse a very attentive reader. 
 We have given the impression  that the theory of electromagnetism and the weak force arise from 
 a gauge theory with the symmetries of a circle and a sphere respectively. 
 In nature, the 
   $W^\pm$ bosons are charged. 
 A charged object is one that is changed under a gauge transformation. 
 This would be impossible 
 if electromagnetism corresponded to a circle completely independent of the weak sphere. 
 In fact, while it is correct that we start with the symmetries of  a circle and a sphere, 
 the symmetry of electromagnetism corresponds to a combination of a rotation on the circle, as well 
 as a rotation on the sphere. Then the $W^\pm$ bosons are charged because when we perform an electromagnetic 
 ``gauge transformation'' we are also performing a rotation on the weak sphere and it is therefore not surprising that
 we can change the $W^\pm$ bosons. 
 Similarly, 
  we will see below  that  the 
electron and the neutrino represent the same particle but rotating in different ways in the 
weak sphere. However, the electron and neutrino have different charges. The neutrino, as
it name indicates, is {\it neutral} or uncharged, while the electron is charged.  But, since electromagnetism 
includes a rotation on the weak sphere, this difference in rotation in the weak sphere translates into a different charge. 
 For this reason the whole theory is called electroweak theory. 
 
This is an important point to get the details right, but we will ignore it in the rest of the discussion. 

\section{Quantum mechanics }   

The system we have described so far, through the economic analogy, gives rise to 
what is normally called a classical field theory. A field is a quantity that is defined at each 
point in spacetime. For example, the price of gold is a field; at each point in spacetime it takes 
a definite value. Similarly, the exchange rates are also fields. For each point, we have one exchange rate
per spacetime dimension, since  the number of neighbors that a country has 
 is proportional to the   dimension of spacetime. 
   
In physics this is not the whole story. The dynamics of these fields is ruled by the laws of 
quantum mechanics. An important feature for these laws is that they are probabilistic. 
One might think that in the vacuum all the fields are zero. However, this is not the case, they take 
random values. 
All we can say is that they are given by a certain  probability distribution.
In the economic analogy, we can say that   the 
values of the exchange rates and the price of gold, are all random. This randomness follows a  very 
precise law, which is encoded in the precise form of the probability distribution.
We will not give the precise formula here (it can be found in the appendix), but we simply note
that it is such that configurations of exchange rates and gold prices become less probable when 
there are greater opportunities to speculate.  We have a precise law for the probability of each configuration, but we
cannot predict with certainty, which of the possible configurations will occur when we look at the system. 

Note that all these fluctuations happen at short distances. If we follow a very big circuit, then we pass
through many countries and their exchange rates all average out. In the vacuum, at long distance they 
average out to zero so that we recover the classical result  where the fields are all zero. 
 
 The probability cost that we have to pay when we set the exchange rates to values leading to larger speculative 
opportunities is also related to the energy cost we discussed above. They are essentially the same. Higher energy 
configurations are less probable.  
In nature,  the particles that carry the weak force are very massive. They weigh around a hundred times the mass of the proton, which is a lot
for an elementary particle.  They are called the $W^+$, $W^-$ and $Z$ bosons. 
  Their large mass explains the weakness of the weak force. 
  This large mass implies that we are very unlikely to produce  fluctuations in the ``weak exchange rates''. 
 Therefore,  a particle that interacts only through the weak force, such as the neutrino,  is very difficult to see. 
  In fact, a few per cent of the energy of the sun comes out in neutrinos\footnote{This shows that the weak force is important for the workings of the sun.}. 
  However, we are totally oblivious to these neutrinos. They simply pass through us day and night and we do not see them. 
   You need very big detectors with 
very sensitive electronics to catch a very tiny fraction of them.

 \subsection{The continuum limit and the Higgs boson} 
 
The mechanism described above does give a mass to the mediators of the weak force, but 
it does not explain why there should be a new physical particle, such as the Higgs boson.
Let us explain this through the economic analogy. 
We can  fix the price of gold to one everywhere by a currency (or gauge) transformation. Once this is done, the only remaining 
variables are the exchange rates. In physics this gives rise to a massive particle (with spin one), but 
to no other particle. 
In classical physics, this theory is perfectly consistent without an extra particle. However, in the quantum mechanical version this is not the case, 
especially in the case of the weak force. 

The quantum mechanical version of the continuum limit, 
 indicated graphically in figure \ref{Continuum}, is very subtle.  
A detailed analysis reveals that a theory without extra fields would not permit the 
weak force mediations to have a mass which stays fixed as we take the spacing between the countries 
to zero. Their mass would go to infinity as we take the continuum limit in the quantum theory. 
For this reason everybody expected that the Large Hadron Collider would discover new particles. 
The simplest possibility was to add  just  one new particle.

In the economic analogy these new particles arise when we have 
 more goods that we can carry between the different countries. 
For example we can also have silver. Silver also has a price in each country, 
again chosen arbitrarily by the inhabitants of each country. 
Now, in each country, the ratio between the price of gold and silver is independent of the currency units, it is gauge invariant.
If one ounce of gold costs 2,000 pesos and one ounce of silver costs 1,000 pesos, then  we can say that gold is twice more expensive than silver. 
If we change currencies to Australes, as in figure \ref{PesosAustrales}, the price of gold is 2 Australes and that of silver is 1 Austral. But gold is 
still twice more expensive than silver. 
  Therefore, in this situation,  
we have a quantity that is defined in each country which is independent of the choice of currency. In physics this corresponds to a field that 
gives rise to a physical particle. This is the field associated to the actual physical Higgs boson particle. Note that in the case that we only had 
gold, we also have a quantity defined in each country, which is the price of gold. However, this quantity depends on the choice of currency. Quantities
that depend on the currency units do not reflect real economic opportunities. In physics they are not observable. In fact, 
 we have seen that through a 
 gauge transformation (or currency transformation) we can set the 
price of gold to one everywhere, so that it disappears as a physical field. 
 However, with both gold and silver we have now a physical field. 
 
  One extra field is the simplest possibility. It predicts one new particle. Indeed this extra particle 
was discovered at the Large Hadron Collider in 2012.

Though we have not attempted to give the history of these concepts, we cannot resist making a few
historical comments. 
 Yang and Mills invented the Yang Mills theory to describe mesons. However,  we currently 
use it to describe weak and strong interactions. This is an example where a good idea
was not useful for its original purpose  but was useful for   other problems.  
Though Pauli's objection was right, it was not a  fatal flaw.  It could be fixed in a relatively simple way. 
The mathematics of  gauge theories was discovered before by mathematicians who know these
structures as fiber bundles. 
Inspired by many previous experiments and partial results, 
the theory of electro-weak interactions was written first by S. Weinberg, with important contributions   
by Glashow and Salam. 
Later experiments confirmed this theory and ruled out competing alternatives. The most important current experiment in particle physics is 
going on at   the Large Hadron Collider in Europe. 

Notice the simplicity of physics relative to economics. In economics there are many things that we can trade, not just gold and silver. 
We can also trade among all the countries at the same time. In physics we can trade only with the neighbors. 
It is worth noting that Weinberg's paper is just three pages long, while here it has taken us many more pages to 
give a still  imprecise explanation. 
This is an example of the power of formulas. It is often said that a picture is worth 
a thousand words. A formula is worth a million pictures!

\begin{figure}[h]
\begin{center}
\includegraphics[scale=.48]{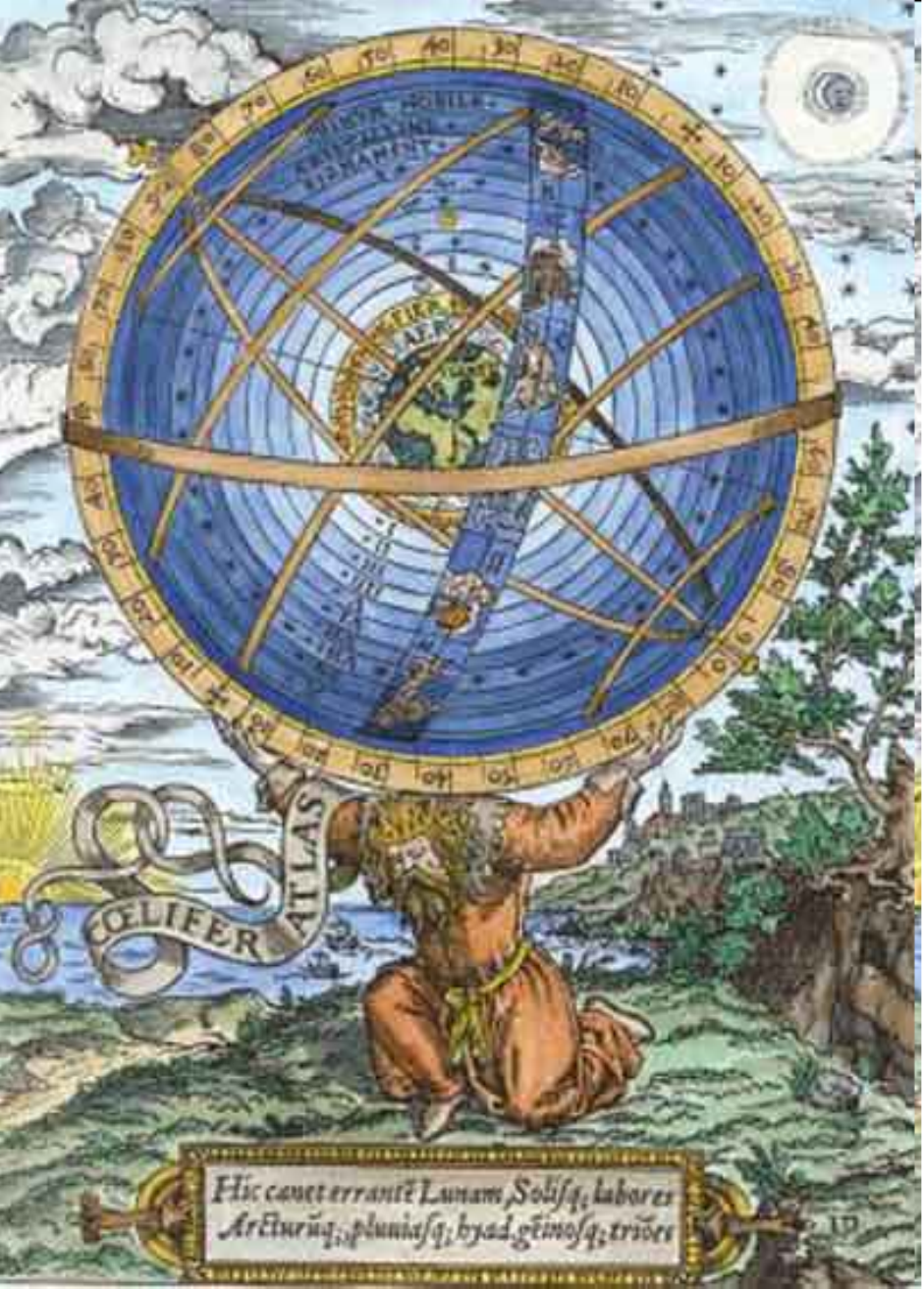} ~~~\includegraphics[scale=.3]{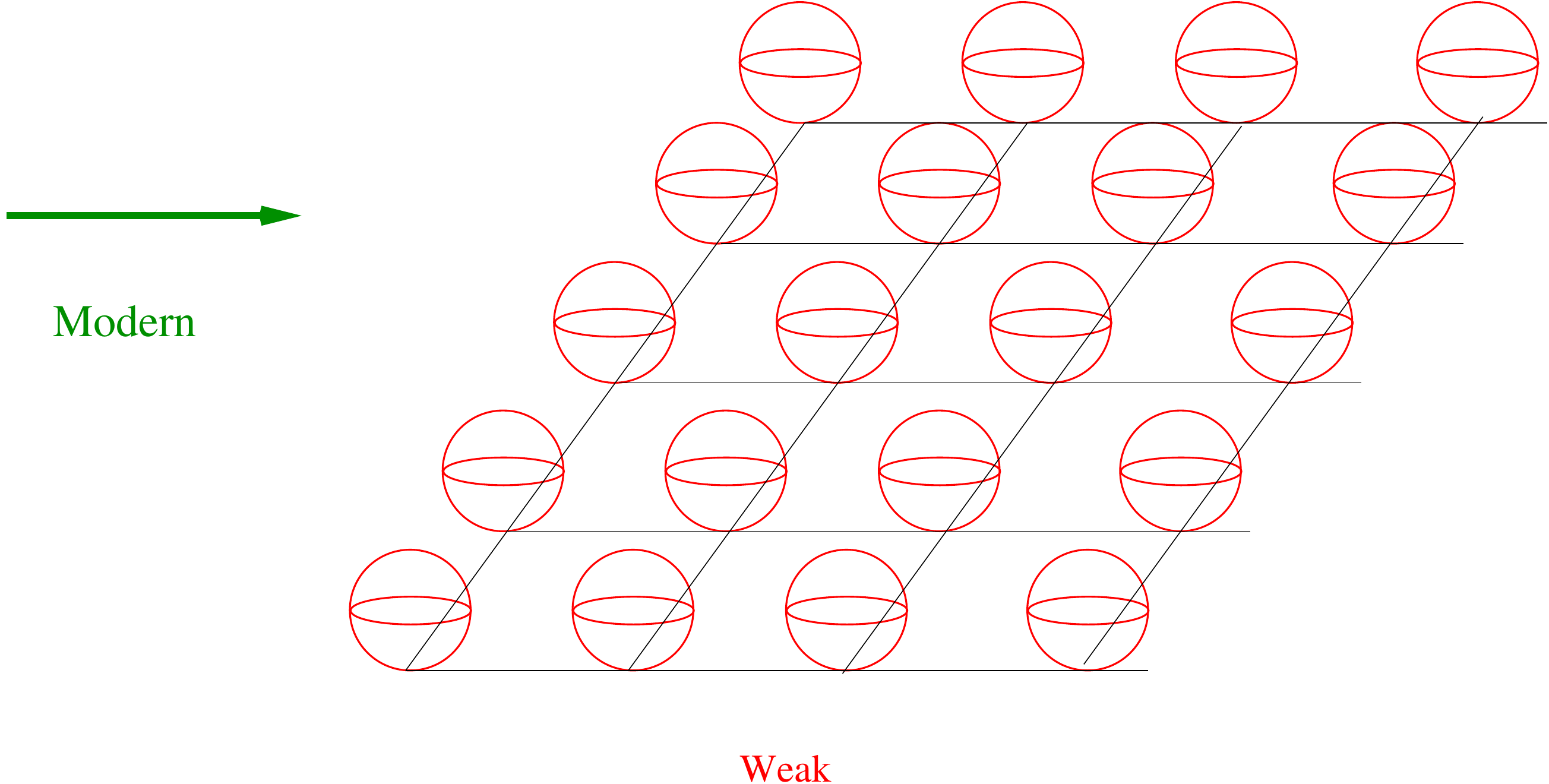}
\caption{ On the left we see Atlas holding the celestial spheres. On the right we see the pattern of spheres whose symmetries determine the 
weak force. The ancients had a few spheres per planet. In some sense, we now have one sphere per spacetime point.   }
\label{AtlasSphere}
\end{center}
\end{figure}

One of the detectors of the LHC is named Atlas. In figure \nref{AtlasSphere} we can see the mythical Atlas holding 
the celestial spheres. Probably when you look at this picture you think: Oh, how naive   these
Greeks were, with all those spheres. How much simpler is the Newtonian description. 
 Now, the modern view of nature  has the symmetries of  a sphere at each point
of spacetime! So the number of spheres has grown a great deal. However, the structure is governed
by a rather simple symmetry. 
 The weak sphere is very simple.  The modern Atlas detector is reaching into  the weak sphere, uncovering its 
secrets\footnote{ My apologies to the 
other LHC detectors which has less mythical  names. From the scientific point of view, they are as important as Atlas! }.

\section{Massses for other particles } 
   
It is often said that the Higgs field gives a mass to all other particles. In fact, from what we said so far, 
we could add masses for the other matter particles, such as the electron, with or without the Higgs. 
The real reason we need the Higgs to give mass to the electron is related to a strange property 
of the weak interactions. To explain this weird feature,  we need to  describe in more detail 
some of the properties of elementary particles. We need to take into account that the electron 
has spin. First we will describe spin, and then describe the weird feature of the weak interactions.

\begin{figure}[h]
\begin{center}
\includegraphics[scale=.48]{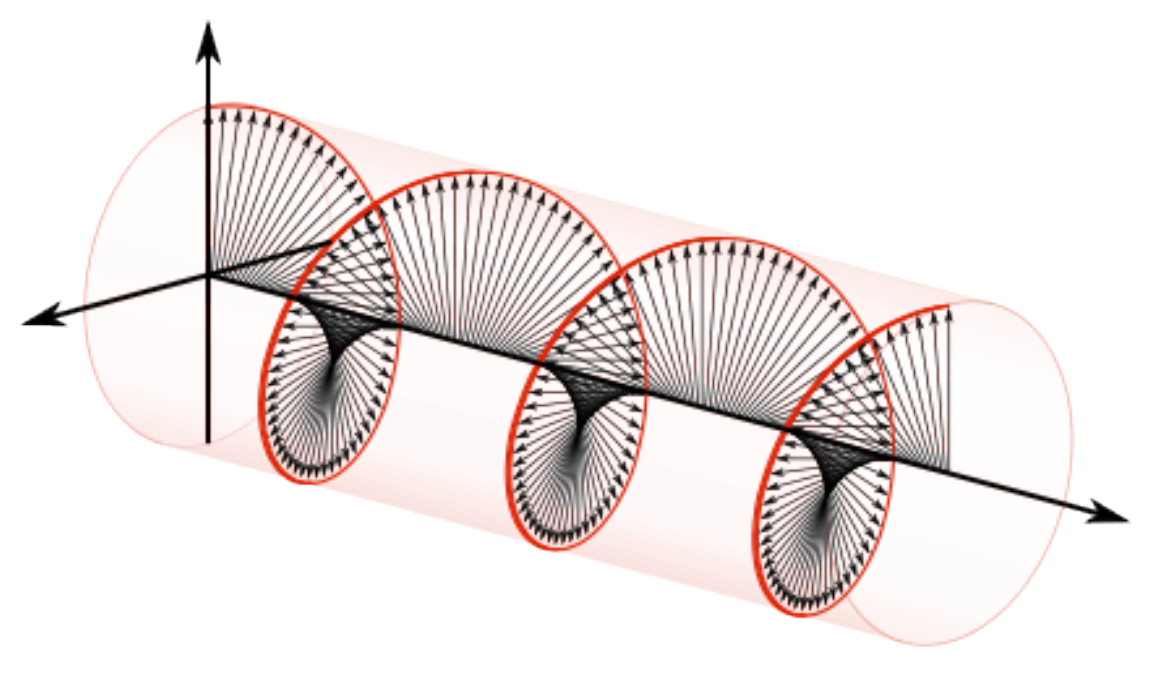} ~~~~ \includegraphics[scale=.45]{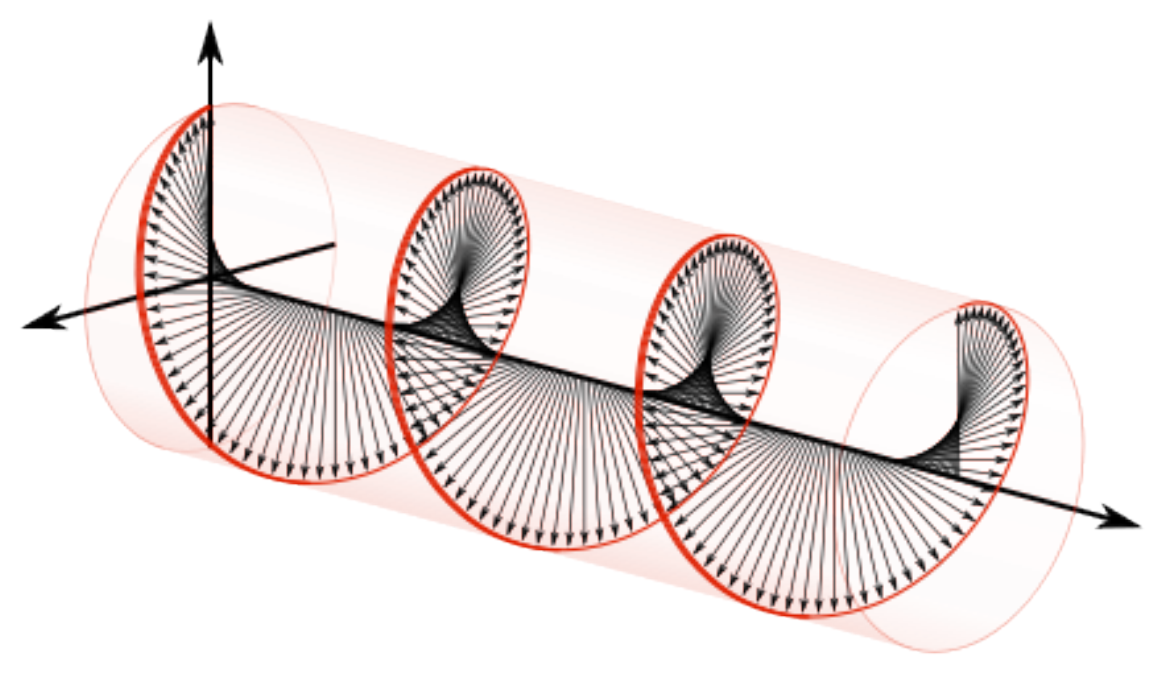}
\caption{  On the left we see a left handed electromagnetic wave. The arrows point along the 
direction of the electric field at one instant in time.
On the right side we see a right handed electromagnetic wave. Now the electric field rotates in the 
opposite direction as the wave propagates. Both waves propagate to the right.    }
\label{LeftandRightHanded}
\end{center}
\end{figure}
You might be familiar with the fact that light can be polarized and that it has two polarization states. 
For example, polarized sunglasses are designed to block the predominantly polarized light  produced by reflection  glare. 
In fact, it is convenient to think in terms of circularly polarized light. 
In figure \ref{LeftandRightHanded} we see a left and a right handed  electromagnetic wave. 
These waves carry angular momentum around the direction of propagation. 
In fact, the photons that form these waves also carry angular momentum. In the economic model, 
the fact that light is polarized is related, to the fact that the exchange rate connects a country to its neighbors. These neighbors 
can be in any of the directions transverse to the wave. In figure \ref{LongShortWavelength} we had a wave going to the right and the exchange 
rates that were changing where the ones going up, a direction  transverse to the wave.

Electrons and neutrinos also share this property with light. They can also be polarized. They also  carry angular momentum, or
some amount of rotation. When a particle is moving we have a preferential direction to define the
angular momentum. This angular momentum can be left or right handed as in the case of light. 
If the particle is at rest we do not have a preferential direction. The angular momentum can point in any direction. All directions  
are related by the rotation symmetry of empty space.  

Something special happens for massless particles. A massless particle is always moving.  If the
angular momentum is along the direction of motion, then in all reference frames it will be along the  
direction of motion. This is not the case for a massive particle. Imagine a massive particle moving up 
in space, with the angular momentum along the direction of motion.  Let us now decide that we are also 
moving along the vertical direction, faster than the particle.  In our moving reference frame the particle looks like it is 
moving down. But its spin is unchanged. So the direction of the spin is now opposite to the direction
of  motion. In other words, by going to a moving reference frame, we have changed the direction
of the velocity relative to the direction of spin.
  The conclusion is that for a massless particle the notion of whether the spin points along the
direction of motion or opposite to it is some characteristic of the particle, independent of the reference 
frame.  Recall that the laws of physics should be the same independent of how we are moving. This is the principle of
relativity. 

\begin{figure}[h]
\begin{center}
\includegraphics[scale=.48]{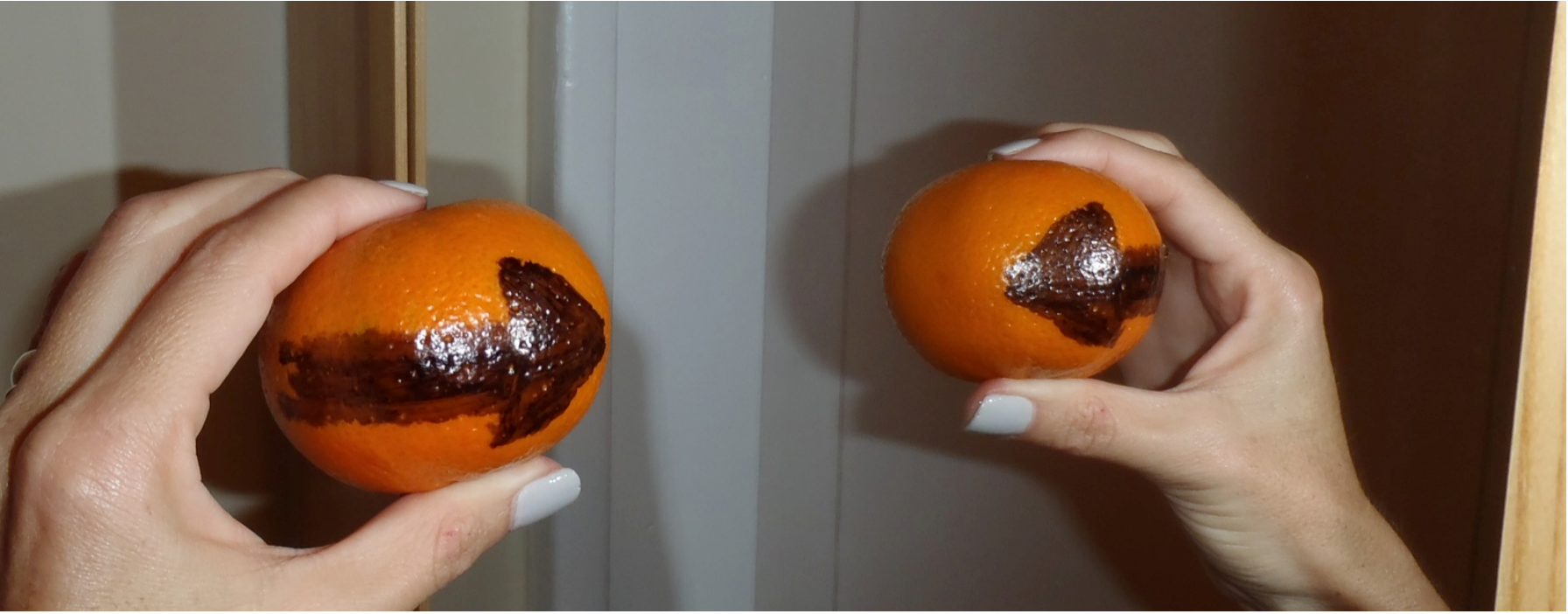}
\caption{ If the tangerine is rotating in the direction of the arrow, then its reflection would 
rotate in the direction of the arrow in the reflection. Therefore,  if the original rotates in one direction, 
the image rotates in the opposite direction.  A tangerine can indeed rotate in any of these two directions. 
  But if we replaced the tangerine by a neutrino moving in the vertical direction,
 then its mirror image would not look
like an allowed particle.   }
\label{MandarinaSmall}
\end{center}
\end{figure}

With these preliminaries, we can now turn our attention to the weird property of the weak interactions. 
The weak interactions treat differently left and right handed particles. The weak interactions affect only 
left handed particles. 
This is a very surprising property, which is possible only  because the weak interactions violate reflection 
symmetry. 
What is reflection symmetry? This is the symmetry under reflections on a mirror. Imagine that
we look  at the world through a mirror. Is the world we see consistent with the laws of physics?
From our everyday experience we would naively think that it should be. It is rather hard to  realize that you are 
looking at the world through a mirror. If you look at written text, then you would know, but this is only  
because we all use the same convention to write. The question is whether the fundamental laws are the 
same or not. 
An important property of reflections in a mirror is that the reflection of a rotating particle appears to 
rotate in the opposite direction. See figure \ref{MandarinaSmall}. Therefore,  if the weak interactions treat
left and right handed particles differently, then the weak interactions can distinguish between the 
real world and its reflection in a mirror. 
We have  emphasized that the laws of physics are based on interesting new symmetries,
 but here we encounter a very simple 
candidate symmetry  that the fundamental laws do not have!
The laws of nature have some unfamiliar symmetries but they lack a very familiar one.

 A left handed electron and the left handed neutrino are basically 
 the same particle but moving in different ways on the weak sphere. The weak force transforms one
into the other.  
 The right handed electron does not feel the weak force. We do not know whether a
right handed neutrino exists or not, it has not been detected yet, and it does not have to exist. 
 This sharp distinction between left and right handed particles is 
 possible if the particles are moving at the speed of light, so that their handedness is an intrinsic 
property.  

\begin{figure}[h]
\begin{center}
\includegraphics[scale=.4]{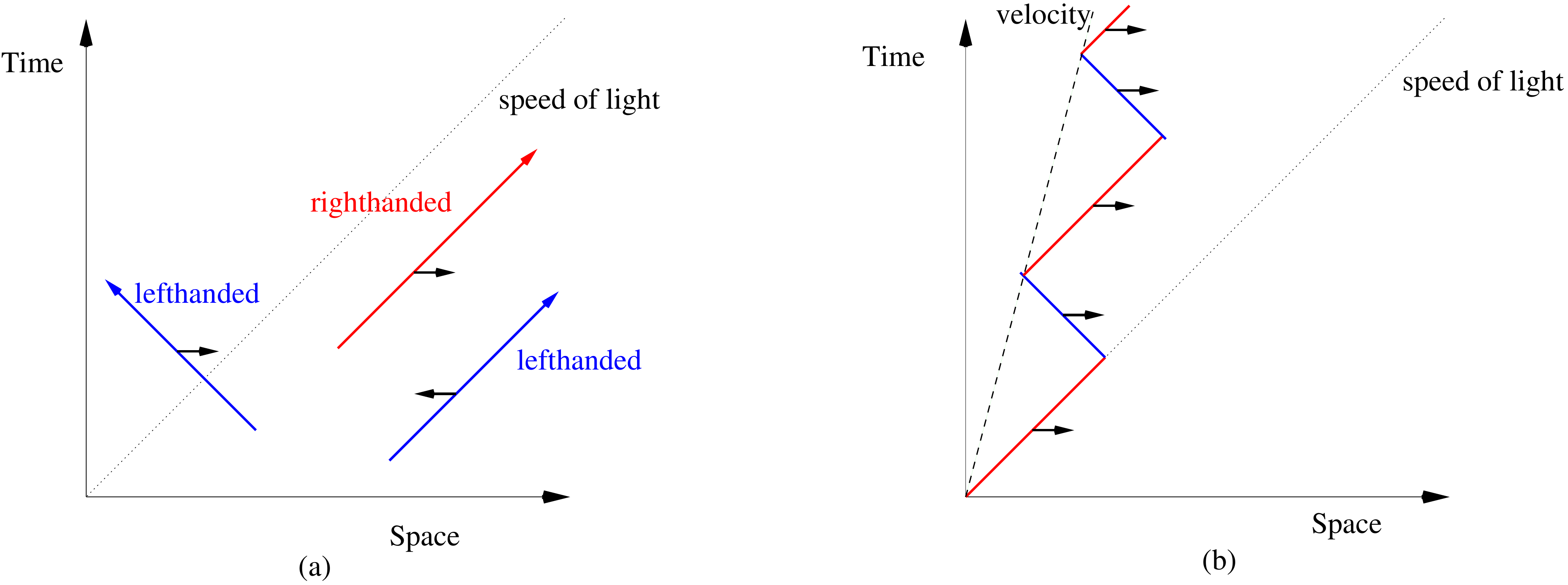} 
\caption{ In (a) we see left and right handed massless electrons. The black arrow is the direction of the spin of the electron. 
Whether it is left or right handed depends on the direction of the spin relative to the direction of motion. So the two blue lines 
describe left handed electrons since the direction of spin is opposite to the direction of motion. 
In (b) we see a massive electron. Some of the time it behaves as a right handed electron and sometimes as  a left handed electron. 
It switches from one to the other thanks to the Higgs field.      }
\label{ElectronMass}
\end{center}
\end{figure}

However, the electron is a massive particle. This is possible only due to its interaction with the Higgs field. 
This is a new interaction that we need to postulate by hand, in order to get the theory to agree with experiment. 
Through this interaction, an electron which is moving at less than the speed of light, can be viewed as a particle with an
 identity crisis. Part of the time it is a left handed electron and part is a right handed electron  moving in the opposite direction. 
In average, it moves at less than the speed of light. The interaction with the Higgs field turns one into the other. 
 See figure \ref{ElectronMass}

The quarks, which are the particles inside the proton or neutron, also get a mass through this mechanism. 
There are further elementary particles similar to the electron, neutrino and quarks, which are more massive. These are  unstable and decay quickly 
via the weak interactions.   All of these get masses 
through the interaction with the Higgs field. These interactions are all introduced by hand in order to fit nature. They do not 
follow from any symmetry principle. Their values range over many orders of magnitude. For example, the top 
quark is about three hundred thousand times heavier than the electron. The neutrinos also get a mass in a similar fashion, though 
the details are a bit more complicated and have not been completely settled yet.  

Despite the fact that  the Higgs gives mass to most elementary particles, most of the mass of 
ordinary objects does not come from   the Higgs field! In fact, most of the mass comes 
from the mass of the protons and neutrons. These are composite objects.  They contain quarks
which are moving very rapidly and most of the mass comes from the energy of this rapid motion.
(Recall $E = m c^2 $.) 
But the Higgs is important for an important  macroscopic property of the real world. 
The Higgs field sets is the size of atoms. The size of atoms is inversely proportional to the mass of
the electron. The lighter the electron the bigger the atom. 
By changing the magnitude of the Higgs field it is possible to make all elementary particles lighter. 
For a person trying to loose weight, it would not be a good idea to try to change the magnitude of
the Higgs field. Ignoring the fact that this would be exceedingly difficult to do, it would not have
the desired effect. The mass would be almost unchanged but  the person's 
size would become much bigger!

Given that the Higgs does all these good things for us, why do we say that the Higgs is ugly?
One reason is that it represents a new force of nature that is not based on a gauge symmetry. 
A more practical one is that most of the unpredictable parameters of the Standard Model are associated
with the interactions with the Higgs field.
These parameters range over many orders of magnitude. By comparison the  strength of 
the three gauge theory forces   are somewhat similar, at high energies. 
 Finally, the  strangest feature of all is the value that sets the scale of the Higgs mass, and as a 
consequence the overall values of all the other masses. 
 It is unclear what physics gives rise to this 
mass scale, and it is a mass scale that is much lower than the other fundamental mass scale in nature
which is the mass scale of gravity, which sets the strength of gravity.  
This can be understood as follows. In the economic model, we have said that we recover the continuum description by 
saying that the distance between the countries is much smaller than the shortest distances we can measure. 
In a world without gravity this distance could be infinitesimally small. In our universe we have gravity. The Einstein
theory of gravity says that spacetime is dynamical. We also expect it to be quantum mechanical. Quantum mechanics 
says that spacetime itself is fluctuating. In the economic model, spacetime is the grid of countries. Having the grid fluctuate means
that countries can exchange neighbors, new countries could appear and disappear, etc. In nature, 
 all of this would be happening at 
a very short distance, a distance set by  the strength of gravity. This turns out to be an exceedingly small distance. 
A distance  
$ 10^{16}$ times smaller than the {\it  shortest } distance  we can see today with our most powerful microscope, which is the Large Hadron Collider. 
The problem with the mass of the Higgs particle 
 is this question of why are the weak interaction phenomena happening at a distance scale so much larger
than this basic gravitational scale?  We do not know.

There are several ideas for understanding these strange features of the Standard Model. 
Many of these ideas postulate the existence of new particles. Perhaps, they will be soon 
 discovered at the Large Hadron Collider. We are waiting expectantly. 

In this exposition we did not talk about the strong interactions. It is the force that holds the quarks together in the proton or
neutron. It is also based on a gauge interaction. 
To answer Pauli's objection,  it uses a  another mechanism which is inherently quantum 
mechanical. 

We should   end by mentioning that we have very strong evidence for the existence of 
a new particle. This evidence comes from astronomical observations which see more matter than 
what can be accounted for with the   particles we already know. This is the so-called ``dark matter''. 
A good candidate for a dark matter particle is a new particle that is subject to the weak interactions.
A so-called WIMP (Weakly Interacting Massive Particle). It is a good candidate because the 
cosmological abundance of these particles would be in the right range to fit observations. 
It is also predicted naturally by theories that try to explain some of the puzzles of the Standard Model. 
It is also possible  that dark matter is not related at all to the weak interactions. 

Hopefully, once  we have a more complete understanding we will find that the 
Higgs field, which we view as an ugly component of the Standard Model, will be transformed into 
a handsome prince. Or at least it will be part of the handsome prince. 
{\it And they will live happily ever after... or at least till the universe decays.}


\be
{\cal T  H E  ~~E N D } \notag
\ee

{\bf Acknowledgments} 

I thank Graham Farmelo for encouraging me to write this up. I  thank N. Arkani Hamed and C. Morgavio for comments on the draft.
I thank Wikipedia for providing some of the pictures. 

 I was funded in part by DOE grant de-sc0009988.

A  video presentation based on  this article can be found in:

  https://www.youtube.com/watch?v=OQF7kkWjVWM

\section{Mathematical appendix:  A more quantitative description of the economic analogy} 

  {\it Warning: This appendix is designed only for people with the right mathematical background. }

Imagine that we have ``countries''   arranged on a $d$ dimensional lattice labeled by points 
$\vec n = ( n_1, n_2, \cdots , n_d)$, where each $n_i$ is an integer number. For our spacetime we would
take $d=4$. 
Each point in the lattice is a country and is labeled by the vector $\vec n$. 
Let us consider the country sitting at the point $\vec n$ and its neighbor in the $i^{th}$ direction, sitting at 
$\vec n + \vec e_i$,  with $\vec e_i=  ( 0,\cdots, 0 ,1 , 0 \cdots 0) $ where the $1$ is at the $i^{th}$ place. 
Here $i$ ranges from one to $d$. The exchange rate between these two countries can be written as 
\be  \notag
 R_{\vec n , i }  = e^{ A_i ( \vec n ) } 
\ee
Here $R_{\vec n , i}$ is the exchange rate and $A_j(n)$ is simply its logarithm, which  we introduce for 
later convenience.  
If the country at point $\vec n$ uses Pesos and the country at point $\vec n + \vec e_j$ uses 
Dollars, then  $R_{\vec n, j}$ tells us how many Dollars you get for one Peso.  

Now a gauge transformation at point $\vec n$ changes the local currency by multiplying it by a factor $f(n)$, which 
we write as 
\be \notag
 f(\vec n) = e^ { \epsilon(\vec n) }
\ee
This changes all the exchange rates connected to this point. 
More explicitly,  under arbitrary currency unit transformations the exchange rates change as 
\be  \label{gaugetr}
 R_{\vec n , i } \to  { 1 \over f_{\vec n}}     f_{\vec n + \vec e_i } R_{\vec n , i } ~,~~~~~~~~~
{\rm or } ~~~~~A_i(\vec n) \to A_i(\vec n) + \epsilon(\vec n + \vec e_i) - \epsilon(\vec n) 
\ee

\begin{figure}[h]
\begin{center}
\includegraphics[scale=.5]{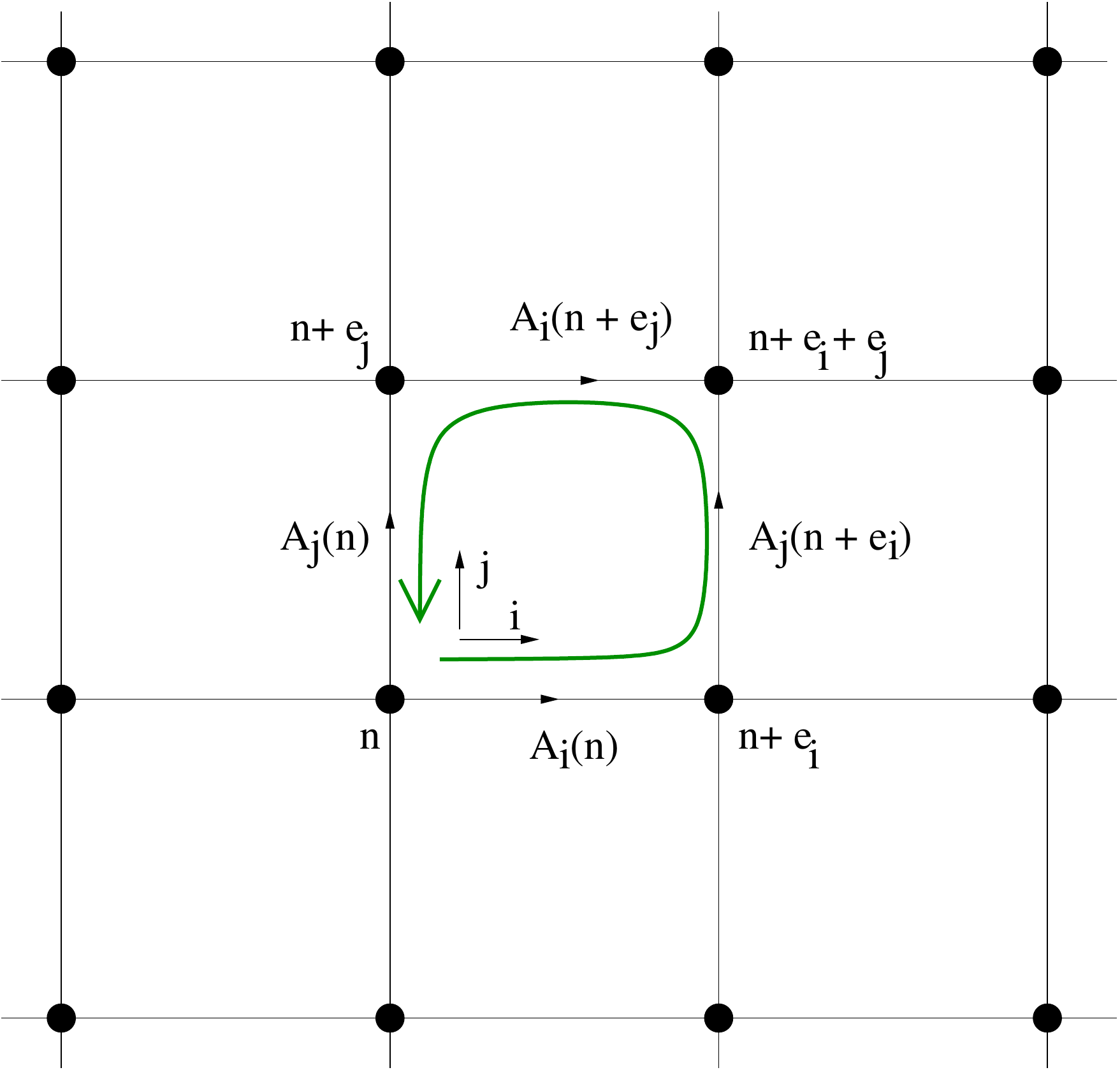}
\caption{ Elementary monetary speculative circuit. We start with some money at 
the country at position $\vec n$. First we move to its neighbor in the 
position $\vec n + \vec e_i$. Then we move to another neighbor at $\vec n + \vec e_i + \vec e_j$. Then to its neighbor at $
\vec n + \vec e_j$. Finally we return to the original country. Following this circuit, and carrying only money, we can 
earn a profit given by the  magnetic flux, \nref{flux}. }
\label{Circuit}
\end{center}
\end{figure}

When we go though an elementary basic square circuit, see figure \ref{Circuit},  the gain  factor is given by 
\bea \notag
   {\rm gain} &=&   R_{\vec n , i} R_{\vec n + \vec e_i , j } { 1 \over R_{\vec n +  \vec e_j  , i } }{ 1 \over 
R_{ \vec n  , j } }  =  e^{F_{ij}(\vec n ) } 
\\ && \notag \\ 
\label{flux} 
F_{ij}(\vec n) &=& A_j(\vec n + \vec e_i) - A_j(\vec n ) -  [ A_i(\vec n + \vec e_j) - A_i(\vec n) ] 
\eea
where we defined the ``magnetic flux''  $F_{ij}$ for  the corresponding elementary  square circuit. 
 Note that the gain factor, or the magnetic flux, is invariant under the change of currency, or gauge transformation, given in 
 equation (\ref{gaugetr}). When the gain factor is less than one you are losing money.

\begin{figure}[h]
\begin{center}
\includegraphics[scale=.7]{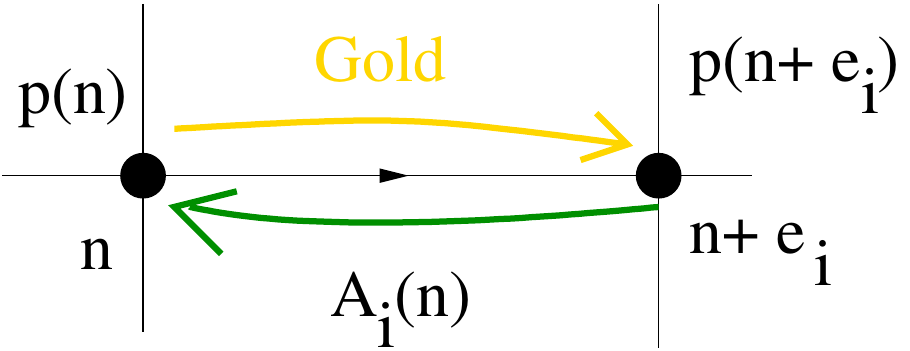}
\caption{ Elementary speculative gold  circuit. We start from the country at position $\vec n$ and buy Gold. We take it to the neighboring country at 
$\vec n + \vec e_j$. We sell it there. We bring back the money to the original country.   Here Gold is yellow and money is green.   }
\label{GoldCircuit}
\end{center}
\end{figure}

We now consider the same but with the addition of gold. Let us write the price of gold as 
\be  
p(\vec n) = e^ {\varphi(\vec n)} 
\ee
Now there is a new opportunity to speculate by taking gold and  bringing back 
money, see figure \ref{GoldCircuit}.  The gain is 
\be \label{goldopportunity}
   {\rm gain} = { p(\vec n + \vec e_i) \over p(\vec n) R_{\vec n, i }  }     = e^{ D_i \varphi(n) }  ~,~~~~~~~ 
D_i \varphi(n) \equiv \varphi(\vec n + \vec e_i) - \varphi( \vec n ) - A_i(\vec n) 
\ee
 where we have defined $D_i \varphi(n)$, 
which in physics is called the ``gauge invariant gradient of the field $\varphi$''. 
We will call this the ``gold gradient". 
It parametrizes  the effective gain of the gold circuit, just as the magnetic flux \nref{flux} was parametrizing
the gain of the monetary circuit.

\subsection{ Quantum mechanical version }

We can define a probabilistic version of the above model by assuming that
 the exchange rates are random.  
They are random variables  drawn from a probability distribution that depends on the magnetic fluxes. 
We further assume that the probability distribution has a local form, so that the probabilities 
of separated  circuits  simply multiply as if they were independent events. The price of gold is 
also a random variable.  There is a  probability for the 
money circuit and also for the gold circuit, see figures \ref{Circuit}, \ref{GoldCircuit}. 

More concretely, the probability for a given set of exchange rates and gold prices  is 
\be \label{probdis}
P[ A,\varphi]  =  \prod_{\vec n,  i , j} \mu( F_{ij}(\vec n) ) \prod_{n , i } \tilde \mu( D_i \varphi(n) ) 
\ee
The product runs over all countries, which are  labeled by $\vec n$.
 For each country we also multiply over   all the elementary money and gold circuits that 
pass through that country. 
Here 
 $\mu(Y)$ and $\tilde \mu(Y)$ are functions which are both peaked at zero. This gives the highest probability to the case where we have
 no opportunity to speculate. We will further assume that these probabilities can be written as 
\be \notag
  \mu(Y) \sim e^{ - Y^2  + \dots } ~,~~~~~~~~\tilde \mu(Y) = e^{ - \sigma^2 Y^2  + \cdots } 
  \ee
  where the dots represent 
higher order terms will not be important in the continuum limit. Here $\sigma$ is just a parameter. 
   In this situation the probability distribution 
  \nref{probdis} simplifies to 
  \be \label{probcuad}
  P[ A , \varphi ] = e^{ - \sum_{n}  \left[ \sum_{i,j} F_{ij}(\vec n)^2 + \sigma^2
\sum_i  ( D_i \varphi(\vec n) )^2 \right] }
 \ee
 
  In physics this gives the actual probability for finding the corresponding magnetic potentials in the vacuum
through the following procedure. 
 We consider
  a four dimensional Euclidean lattice of this form.
 We focus on the particular set of exchange rates and gold prices 
 sitting at $n_4 =0$. This is a particular three dimensional 
  sublattice of the original lattice of countries. 
 This surface can be interpreted as a discrete version of 
   physical space at some instant of time, say at $t=0$. 
  The probabilities for the links at $n_4=0$ in the above Euclidean model are essentially the same as the full quantum mechanical probabilities 
  for measuring the corresponding values of the magnetic potentials at an instant of time, say $t=0$, in the vacuum. 
  
  We say ``essentially'' because we  still need to take the continuum limit,
 indicated in figure \ref{Continuum}. 
  This can  be done as follows. We imagine that each point in space is given by 
  $\vec x = a \vec n$ where $a$ is a very small number that goes to zero and $\vec n$  goes to infinity so that $\vec x $ stays fixed. 
  In this situation it is convenient to introduce a new magnetic potential , $ {\cal A}_i(\vec x) $,
that will stay fixed as we take the continuum limit  
  \be  \notag
  A_{i}(\vec n) \to a  { \cal A}_i(\vec x )     ~,~~~~~~~~~~~  \varphi( \vec n) \to \phi(\vec x) 
  \ee
  We also have 
  \bea \notag
  && F_{ij}(\vec n)  \to   a^2 {\cal F}_{ij}(x) ~,~~~~~~~{\cal F}_{ij}(\vec x) \equiv  { \partial {\cal A}_j \over \partial x^i } -  { \partial {\cal A}_i\over \partial x^j }
  \\ \notag
  && D_i \varphi(\vec n) \to a  D_i \phi(x)   = a \left[  { \partial \phi(\vec x) \over \partial x^i } - { \cal A}_i(\vec x ) \right]
  \\ \label{probcont}
  && P[ {\cal A }, \varphi ] = e^{ -  \int dx^d  \left[ \sum_{i,j} {\cal F}_{ij}^2  +m^2 \sum_i ( D_i \phi )^2 \right] } ~~~~~~~~~ m^2 \equiv { \sigma^2 \over a^2 } 
  \eea
 
 Now, this is the full story with a massive spin one field. This is the real physical theory. As you see it is not that complicated. 
 We have obtained it from a relatively simple probabilistic economic model. 
 What is somewhat complicated is to relate this description to actual physical measurements. 
 The final parameter $m$ is the mass of the spin one particle.  We recover ordinary electromagnetism by setting $m=0$.  
 
 As we have said we can get the quantum mechanical probabilities for the vacuum from \nref{probcont} by fixing the ${\cal A}_i(x)$ fields 
 at one instant of
 Euclidean time $x_d=0$ and then integrating out over the fields at all other times. 
  The description of quantum mechanical processes involving time, such as observations at different values of ordinary time, is more 
  complicated and would require a longer discussion of quantum mechanics. 
But conceptually, it is basically the same as what we have done so far. 
The rest of the forces of particle physics, the weak force and the strong force can be incorporated through a similar description. 
Some  details  are somewhat complicated due to the need to incorporate variables that are anticommuting ``numbers'' to describe electrons, neutrinos or quarks. 

Finally, Maxwell's  equations can be simply derived from the
 condition that the exchange rate variables ${\cal A}_i(x)$ are such that they 
minimize the probability.  Even though the fundamental description is random 
we can try to find the average exchange rates and gold prices  that maximizes the probability  \nref{probcuad} or \nref{probcont}. 
These are obtained by taking the derivative of the exponent in \nref{probcuad} with respect to $A_i(\vec n)$ and $\varphi(\vec n)$ and setting these derivatives
to zero. This gives a discrete version of the classical equations of motion for a massive field. Doing this directly in the continuum we get the 
standard continuum euclidean equations 
\be \label{massmaxcont} 
\sum_{j=1}^d  { \partial {\cal F}_{ij} \over \partial x^j} - m^2  D_i \phi =0 ~,~~~~~~~ {\partial D_i \phi \over \partial x^i } =0 
\ee

\subsection{ Another amusing way to obtain the classical equation in the economic model} 

Here we derive the classical equations of motion directly in the original economic model without
going through the probabilistic interpretation.  
 If we started with an arbitrary configuration of exchange rates and gold prices, we   expect that
speculators would  start moving around and earning
money in the process. 
 Let us focus on one of the banks that sits between two neighboring countries. 
 Let us say  the currency of one is Pesos and the other is Dollars. If there are more speculators
 trying to buy Dollars than there are trying to buy Pesos, then, in the real world, the bank  
would try to change the exchange rate so that there is no imbalance. 

In order to model this situation, we assume that 
   the magnetic flux \nref{flux} or gold price gradient \nref{goldopportunity} are both 
very small. We also assume that the exchange rates are very close to one to one, and that
gold prices are all very similar.  In this world, the 
opportunities to speculate are very small.  
We also  assume that speculators follow only the two elementary 
circuits  in figures \ref{Circuit}, \ref{GoldCircuit}. Of course, they can start from any country. 
     We also make the important assumption  the total amount of money carried by 
 speculators following each circuit 
 is proportional to the gain of each circuit
\be \notag
{\rm money~ carried ~ by ~speculators}  = ({\rm constant} ) F_{ij}   ~~~~~~~{\rm or}~~~~ = ({\rm constant}) D_i \varphi
\ee
 This statement is a bit ambiguous because we did not specify the currency. However, we have
assumed that all exchange rates are close to one to one, therefore the units do not matter for small 
values of these exchange rates. We also restrict the gauge transformation parameters $\epsilon(\vec n)$ to 
be small. 
 When we say that the
 speculators carry an amount of money proportional to the flux, this money can be specified in 
 any of the currencies 
 on the circuit. It does not make a difference when we work to first order in the fluxes. 
 As the speculators go around the circuit, they will make a small profit
 proportional to the magnetic field $F_{ij}(\vec n)$.  As a consequence of our assumptions, this 
  is small compared to the 
 initial amount. In other words, it is a very small percentage.

 In this situation the net amount of money flowing through a given bank, say the bank that 
 sits between the countries at point $\vec n $ and $\vec n + \vec e_i$,  is 
  proportional to the number of speculators crossing between these countries. Of course, 
   speculators 
  crossing from $\vec n + \vec e_i$ to $\vec n$ count with a minus sign. We want this net flux of 
  money to be zero so that the bank does not run out of either of the two currencies. 
Taking into account both the monetary circuit and the gold circuit this imposes the condition
 \be \label{maxmat} 
\sum_{j=1}^d  (  F_{i,j}(\vec n) - F_{i,j}(\vec n - \vec e_j) )  - D_i \varphi(\vec n)  =0
\ee
Note that all the elementary circuits that share the link going from $\vec n $ to $\vec n + \vec e_i$ appear in this sum. 

Similarly we assume that the price of gold at each country adjusts so that there is no net 
gold inflow or outflow.  Otherwise the inhabitants of this country would change their price of gold. 
 This implies
\be \label{maxmass}
\sum_{i=1}^d  D_i \varphi(\vec n) - D_i\varphi(\vec n - \vec e_i )  =0
\ee
This is summing   over the contributions from speculators following the 
elementary Gold circuits along all the bridges connected to a given country. 

These equations,  \nref{maxmass} and \nref{maxmat}, 
  become \nref{massmaxcont} in the continuum limit. These are the magnetic part of
Maxwell's equations in a time independent case. 
 In this model, the Maxwell equations arise from the behavior of speculators that are 
present at short distances. In physics, there are theories where the equations of electromagnetism 
arise from the presence of a large number of very massive charged fields. At long distances the effect
of such particles is to induce the equations of electromagnetism.

\subsection{ Introducing time } 

\begin{figure}[h]
\begin{center}
\includegraphics[scale=.5]{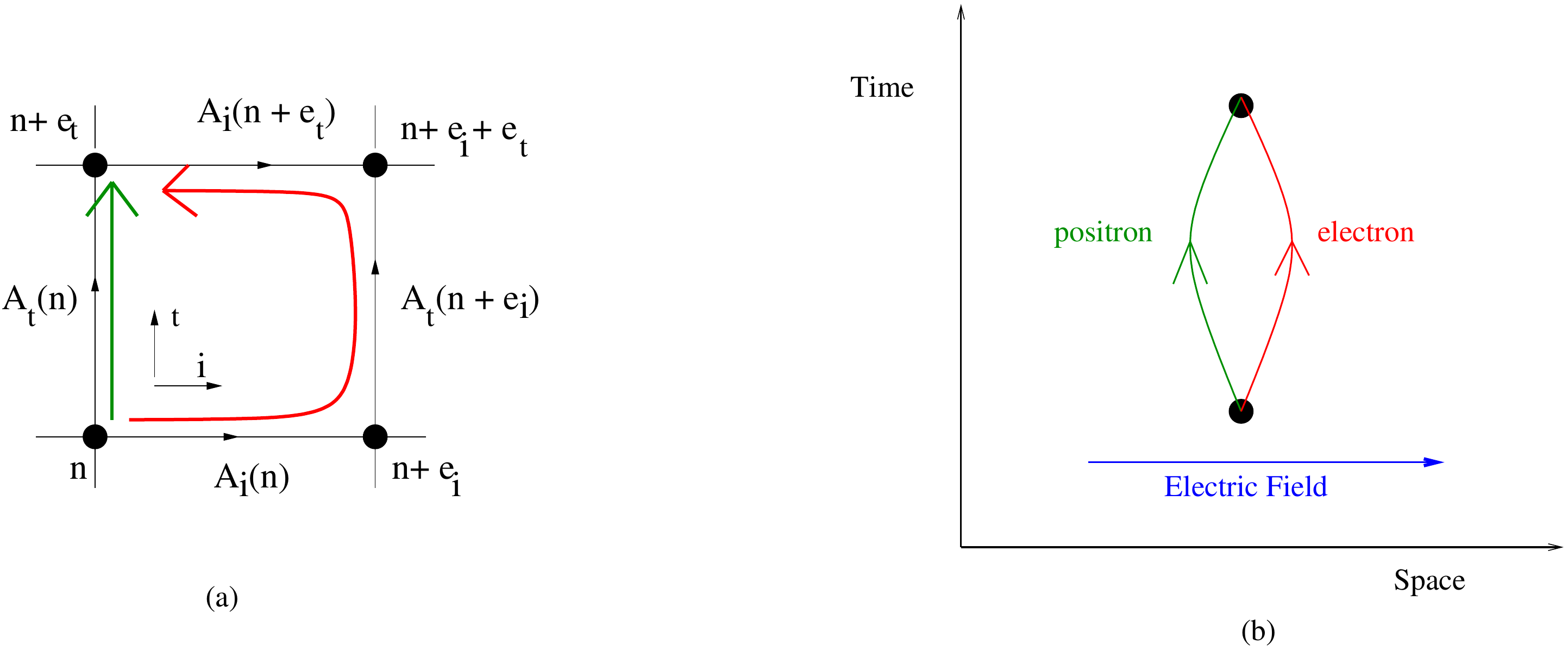}
\caption{ (a) Economic model when we include the time direction. The vertical direction is time. 
 Here green is money and red is debt. Following this circuit 
there can be gain if $F_{ti}$ is positive. (b) Corresponding process in real electromagnetism. Here some external agent creates a positron and 
an electron moving in opposite directions. The electric field curves their trajectories and makes them meet again. In this process there is a 
net ``gain''. When particles try to take advantage of this gain, they end up moving as if they were accelerated by the electric field.    }
\label{TimeCircuit}
\end{center}
\end{figure}

So far, we have been discussing the model in Euclidean space, ignoring the time direction, or
taking it to be equivalent to the space dimensions. 
Let us now include it more properly. 
  We can think of one of the directions in our lattice as a time direction, say it is the $d^{th}$ 
dimension, and label it by the index $t$. The exchange rate in the time direction is simply saying that if you 
have some amount of money 
at some instant in time, then at the next instant your money will be converted 
  to 
$e^{ A_t(\vec n) } $ times your original amount. Each instant in time has its own currency. 
Equivalently, you can think of $A_t(\vec n)$ as the central bank interest rate of the corresponding 
country. And you are {\it required} to deposit your money in the central bank. 
 Of course there can be opportunities to speculate by going around 
circuits that have one side along the time direction. You might say that it is impossible to 
travel backwards in time. However, you can do the following. See figure \ref{TimeCircuit}. 
You  
make an arrangement with another speculator. You borrow some money and  give it to  him.
 Now you have debt and he has money. He stays in the original
country and you move to the neighboring country at the initial instant of time, you wait there till the next instant (depositing the money   in the corresponding central bank), 
and then you return to the original 
country. If you did this properly, and if $F_{t i}$ is non-zero,  he would end up with more money that your
 debt.  You can cancel the debt and share the profits with your friend.  
In physics, 
this would be analogous to a situation where you create an electron an positron pair at some point in spacetime with some initial velocities so that
they run away from each other. Then the electric field pushes them back together at a later instant in time.

Let us now derive Maxwell's equations. 
Let us consider the case with no gold, so that  
speculators  can only carry money between different countries. 
 We  assume that we start with a spatial lattice as before. For our spacetime we would start from a three dimensional 
 lattice of countries.  Let us take time to be continuous, and think of $A_t$ as the central bank exchange rate. 
For simplicity, let us choose the currency of each country so that we can set $A_t=0$. This is like making
a continuous adjustment of the currency units.  
 As before, we   assume that the amount of money that speculators carry  per unit time
   around the elementary spatial  circuits is proportional to the  
  magnetic flux of the spatial circuit. If the net flux of money at a bank is non-zero, then the bank  starts accumulating one of the two currencies. 
  It will have an imbalance.   The imbalance at the particular bank sitting between the countries at $\vec n$ and $\vec n + \vec e_i$ is 
changing as 
  \be  \label{imbalance}
 { d  I_i(\vec n) \over d t }  =- \sum_{j=1}^{d-1}     [  F_{i,j}(\vec n) - F_{i,j}(\vec n - \vec e_j) ] 
\ee
Here $I_i(\vec n)$ is the total imbalance of the bank. It is the difference between the amount of currency of the country
 at $\vec n + \vec e_i$ minus the total amount currency of the
country at $\vec n$ that the bank has. 
Now we add a new rule. We assume that when the bank sees an imbalance $I_i(\vec n)$ it  starts changing the exchange rate with a 
speed which is proportional to the imbalance
\be
 { d  A_i(\vec n) \over d t }  =  I_i(\vec n)  \label{rateofchange}
\ee
By taking a time  derivative of \nref{rateofchange} and using \nref{imbalance} we obtain 
\be
 { d^2  A_i(\vec n) \over d t^2 }  = - \sum_{j=1}^{d-1}     [  F_{i,j}(\vec n) - F_{i,j}(\vec n - \vec e_j) ] 
\ee
which becomes of the Maxwell's equations  in the continuum limit.
This is a wave equation which predicts the electromagnetic waves of figure \ref{LeftandRightHanded}.
 The other equation, which is Gauss's law, says 
\be \label{gauss} 
 \sum_i  { d A_i(\vec n) \over d t } - { d A_i(\vec n - \vec e_i) \over dt } =0 
 \ee
 This can be derived  by assuming that there are 
speculators going around the time circuits, see figure \nref{TimeCircuit}.
 These speculators also carry an amount of money proportional to the gain on the circuit. The
gain is proportional to ${ d A_i \over d t }$. 
Demanding that there is no net amount of money deposited at each of the countries central banks
imposes \nref{gauss}.  We can restore a generic value of $A_t$ by replacing ${d A_i(\vec n) \over d t} \to 
F_{ti} = {d A_i(\vec n) \over d t}  - [A_t (\vec n + \vec e_i) -A_t(\vec n) ] $ in the above equations. 
 
 With similar assumptions we get the equation for a massive field when we include gold. 
With gold one needs to assume that the price of gold obeys an equation of the form 
${d p(\vec n) \over d t } = - G(n)$ where $G(n)$ is the amount of gold at each country, etc.

\end{document}